\documentclass[usenatbib]{mn2e}
\usepackage{apjfonts}
\usepackage{epsfig}
\usepackage{amssymb}
\usepackage{amsmath}
\usepackage{fixltx2e} 

\newcommand{\scaleup}{}

\newcommand{\plotter}{\plotone}

\newcommand{\mdot}{\dot{m}}
\newcommand{\etal}{et al.}

\newcommand{\msun}{M_{\sun}}

\newcommand{\lbol}{L_{\rm bol}}
\newcommand{\ergsec}{{\rm erg\,s^{-1}}}

\newcommand\plotone[1]
 {\centering \leavevmode \includegraphics[width={0.99\columnwidth}]{#1}}
\newcommand{\plotside}[1]
 {\centering \leavevmode \includegraphics[width={0.95\textwidth}]{#1}}
\newcommand{\acknowledgements}{\begin{small}\section*{Acknowledgments}\end{small}}
\newcommand\altaffilmark[1]{$^{#1}$}
\newcommand\altaffiltext[1]{$^{#1}$}
\title[Seyferts Across Observed Wavelengths]{Are Most Low-Luminosity AGN Really Obscured?}

\author[Hopkins \etal]{
Philip F. Hopkins\altaffilmark{1},\thanks{E-mail:phopkins@astro.berkeley.edu} 
Ryan Hickox\altaffilmark{2},
Eliot Quataert\altaffilmark{1},
\&\ Lars Hernquist\altaffilmark{2}
\vspace*{6pt} \\
\altaffiltext{1}{Department of Astronomy and Theoretical Astrophysics Center, University of California Berkeley, 601 Campbell Hall, Berkeley, CA 94720} \\
\altaffiltext{2}{Harvard-Smithsonian Center for Astrophysics, 
60 Garden Street, Cambridge, MA 02138}}

\date{Submitted to MNRAS, January 18, 2009}

\begin{document}

\maketitle

\label{firstpage}

\begin{abstract}
  At low Eddington ratios ($\dot{m}$), two effects make it more
  difficult to detect certain AGN given a particular set of selection
  methods.  First, even allowing for fixed accretion physics, at low
  $\dot m$ AGN become less luminous relative to their hosts, diluting
  their emission; the magnitude of the dilution depends on host
  properties and, therefore, on luminosity and redshift. Second,
  low-$\dot{m}$ systems are expected and observed to transition to a
  radiatively inefficient state, which changes the SED shape and
  dramatically decreases the luminosity at optical through infrared
  wavelengths. The effects of dilution are unavoidable, while the
  precise changes in accretion physics at low $\dot m$ are somewhat
  uncertain, but potentially very important for our understanding of
  AGN. These effects will have different implications for samples with
  different selection criteria, and generically lead to differences in
  the AGN populations recovered in observed samples, even at fixed
  bolometric luminosity and after correction for obscuration.
  Although the true Eddington ratio distribution may depend strongly
  on mass/luminosity, this will be seen only in surveys robust to
  dilution and radiative inefficiency, such as X-ray or narrow-line
  samples; by contrast, selection effects imply that AGN in optical
  samples will have uniformly high Eddington ratios, with little
  dependence on luminosity, even at low $L_{\rm bol}$ where the median
  ``true'' $\mdot\lesssim0.01$.  These same selection effects also
  imply that different selection criteria pick out systems with
  different hosts: as a result, the clustering of low-luminosity
  optical/infrared sources will be weaker than that of X-ray sources,
  and optical/IR Seyferts will reside in more disk-dominated galaxies,
  while X-ray selected Seyferts will be preferentially in early-type
  systems. Taken together, these effects can naturally explain
  long-standing, apparently contradictory claims in the literature
  regarding AGN Eddington ratio distributions, host populations, and
  clustering.  Finally, we show that if current observed Eddington
  ratio distributions are correct, a large fraction of low-luminosity
  AGN currently classified as ``obscured'' are in fact radiatively
  diluted and/or radiatively inefficient, not obscured by gas or
  dust. This is equally true if X-ray hardness is used as a proxy for
  obscuration, since radiatively inefficient SEDs near $\dot m \sim
  0.01$ are characteristically X-ray hard. These effects can explain
  most of the claimed luminosity/redshift dependence in the
  ``obscured'' AGN population, with the true obscured fraction as low
  as $\sim20\%$.
\end{abstract}


\section{Introduction}
\label{sec:intro}

At the highest luminosities, it is generally accepted that active
galactic nuclei (AGN) must lie within a relatively narrow range of
black hole masses and Eddington ratios -- after all, a quasar with
$\lbol \sim 10^{47}\,\ergsec$ requires a $\gtrsim 10^{9}\,\msun$ black
hole (BH); and since this is already on the exponential tail of the
black hole mass function, there are vanishingly few
$\sim10^{11}\,\msun$ BHs that could give a similar luminosity at low
($\sim0.01$) Eddington ratios. At low luminosity, however, there is an
obvious degeneracy between BH mass and Eddington ratio -- a Seyfert
($\lbol\lesssim 3\times10^{45}\,\ergsec$) could be a low-mass BH at
high Eddington ratio, or a high-mass BH at low Eddington
ratio. Moreover, although there are more low-mass BHs by number
density, the duty cycle of low-Eddington ratio activity is higher in
many physical models \citep[see
e.g.][]{dimatteo:cosmo.bh.growth.sim.1,dimatteo:cosmo.bhs,granato:sam,
hopkins:lifetimes.letter,lapi:qlf.sam,sijacki:radio},
so the two may
contribute comparably or in different proportions as a function of
luminosity or redshift.

Although it is often implicitly assumed that Seyfert samples selected 
in the same bolometric (obscuration-corrected) luminosity range 
are reasonably homogeneous, 
there are at least two effects -- one observational and one 
physical -- that could introduce dramatic differences depending on the wavelengths 
of identification and the methods used to identify AGN.

Observationally, even if accretion were self-similar, i.e.\ there were
no change in SED with $\mdot$, many samples implicitly require that the ratio of
the AGN optical continuum to host luminosity be above a certain
threshold for a galaxy to be identified as an AGN. This criterion
varies between different samples and wavelengths, both because of
physical differences in the SED shapes of AGN and galaxies, and
because of various survey depths and selection methods. Most limited
are surveys which select on the basis of ``typical'' quasar colors
dominated by the optical/UV continuum: this implicitly requires that
the AGN continuum in the bands of interest be at least comparable to,
if not greater than, that of the galaxy.

But even a complete spectroscopic survey looking for Type 1 systems
(i.e.\ the identification of broad lines above some host continuum)
implicitly requires an AGN continuum luminosity greater
than $\sim0.1$ times that of the host \citep[or else the lower-level
broad regions of the lines can disappear into the host continuum; see
e.g.][]{vandenberk:qso.spectral.decomposition,sanchez:qso.host.colors,
  jahnke:qso.spectral.decomposition}.  This implicitly sets an
Eddington ratio cut as a function of host galaxy properties, 
where the Eddington ratio $\mdot$ (the dimensionless mass 
accretion rate) is defined as 
\begin{equation}
\mdot \equiv \frac{\dot{M}}{\dot{M}_{\rm Edd}} = \frac{\dot{M}}{2.4\,\msun\,{\rm yr^{-1}}\,
(M_{\rm BH}/10^{8}\,\msun)}
\end{equation}
(where the last equality comes from assuming a standard radiative
efficiency of $\epsilon_{r}\sim0.1$).\footnote{We also define the
  dimensionless Eddington-scaled luminosity, $\lambda$, in terms of
  the bolometric luminosity 
  \begin{equation}
\lambda \equiv \frac{L_{\rm bol}}{L_{\rm Edd}} = \frac{L_{\rm bol}}{1.3\times10^{46}\,{\rm erg\,s^{-1}}\,
(M_{\rm BH}/10^{8}\,\msun)}.
\end{equation}
This is equivalent to $\mdot$ when the radiative efficiency is
constant, i.e.\ for $\mdot$ above the critical value $\sim0.01$,
$\lambda=\mdot$. At lower $\mdot$, $\lambda$ may drop much more
rapidly than $\mdot$, if sources become 
radiatively inefficient (and $\lambda$ inferred from different bands will
be different).} 

To assess the effects of dilution quantitatively, assume that a system
lies on the \citet{magorrian} relation between BH mass and host bulge
stellar mass $M_{\rm BH}\approx \mu\,M_{\ast,\,\rm bul}$ \citep[where
$\mu\approx0.001$;][]{haringrix}.  The total bulge mass is related to
the total stellar mass by the bulge-to-total stellar mass ratio
$B/T\equiv M_{\ast,\, \rm bul}/M_{\ast}$, and the total host galaxy
luminosity will be correspondingly given by the stellar mass-to-light
ratio in the observed band, $(M_{\ast}/L)_{\rm host}$. The total host
luminosity in the observed band is therefore $M_{\rm BH} /
[\mu\,(M_{\ast}/L)_{\rm host}\times B/T]$.  Assuming for simplicity
that the radiative efficiency is $\sim 0.1$, so that $\lambda =
\mdot$, the AGN bolometric luminosity is simply $L=\mdot\,L_{\rm
  Edd}=\mdot\,(3.3\times10^{4}\,L_{\sun})\,(M_{\rm BH}/M_{\sun})$.  If
the AGN bolometric correction to the observed band is $C_{\lambda} =
L_{\rm band}/\lbol$, then the ratio of the AGN luminosity in the
observed band, to the host galaxy luminosity in the same band, will be
proportional to the Eddington ratio and given by $\frac{L_{\rm
    AGN}}{L_{\rm host}} \approx
33\,\mdot\,C_{\lambda}\,B/T\,(M_{\ast}/L)_{\rm host}$ (with
$(M_{\ast}/L)_{\rm host}$ in solar units).  For the $B$-band, with
typical parameters $C_{\rm \lambda}\approx1/12$ and $(M_{\ast}/L)_{\rm
  host}\sim 0.5-3\,M_{\sun}/L_{\sun}$ \citep[for bolometric
corrections, see][]{richards:seds,hopkins:bol.qlf}, this translates to
\begin{equation}
\frac{L_{\rm AGN}}{L_{\rm host}}\approx \frac{\mdot}{0.1}\ 
\frac{(M_{\ast}/L_{B})_{\rm host}}{3\,(M_{\sun}/L_{\sun})}\ 
\frac{B}{T}.
\label{eqn:dilution}
\end{equation}
Given typical mass-to-light ratios the AGN will only be visible in
color selected samples or identifiable as a Type 1 system by most
survey criteria for high accretion rates $\mdot \gtrsim
0.01-0.1$. Note also the scaling with the host mass-to-light ratio --
a younger or more star-forming galaxy is brighter, masking more of the
AGN continuum emission, yielding a potential bias against such systems
in certain samples of optical/IR-selected AGN
\citep{jahnke:qso.host.sf,jahnke:qso.host.review}.  By contrast, in
the X-rays galaxies have much higher $M/L$ -- a system with an X-ray
luminosity $\sim10^{43}\,\ergsec$ will be well above the luminosity of
even a strongly star-forming host.

Physically, at low Eddington ratio, models predict a transition from a
standard, optically thick, UV-bright thin disk to a radiatively
inefficient, optically thin accretion flow \citep[see e.g.][and
references therein]{NY94, esin:truncated.thin.disk.xrb,
  xbongs,quataert:adaf.spectral.models}, as seen in X-ray binaries
\citep{mcclintock:xrb.review}; many observations of both AGN and X-ray
binaries support this scenario \citep{nym96,meier:jets.in.adaf,
  ho:radio.vs.mdot,maccarone:agn.riaf.connection,
  marchesini:low.mdot.sample,jester:riaf.test,cao:riaf.constraints}.

In radiatively inefficient accretion flows (RIAFs), the thin disk is
assumed to evaporate and/or move to much larger radii, and the broad
line region and most optical/UV emission declines, \footnote{Note
  that, in realistic models, the broad line region does not simply
  vanish instantly, but rather moves outwards with the thin disk and
  so declines with the optical/IR luminosity of the AGN. Of course,
  this decline, relative to the host, will still yield a system that
  will not appear in traditional broad-line samples.  Observationally,
  the presence of some weak broad lines, even in very low accretion
  rate systems (many of which show continuum or polarization evidence
  of being in a RIAF/thick disk state inwards of the thin disk) has
  been noted down to accretion rates $\mdot\sim0.001$ \citep[see
  e.g.][]{ho:llagn.seds,moran:hblr.seyferts,tran:hblr.2,
    hao:local.lf,hao:local.lf.selection}.  } while a hard X-ray
spectrum from the thick disk remains. In such a state, the system
could still appear in X-ray samples with nearly the same hard X-ray
luminosity as it would have if the radiatively efficient solution were
extrapolated to the same physical accretion rate, but the system would
disappear from optical samples looking for optical/UV light above the
host galaxy continuum or for broad-line emission.  The systems would
likely, however, still appear in {\em narrow-line} AGN samples, since
the central hard X-ray source remains a significant source of
photoionization (exactly how the narrow-line properties scale for an
incident RIAF spectrum is not certain; a detailed investigation of
these scalings will be the subject of future work). Indeed,
\citet{kewley:agn.host.sf} argue that the distinction between Seyferts
and LINERs largely owes to Eddington ratio, with LINERs being lower
Eddington ratio systems with a harder photoionizing spectrum,
consistent with our discussion here.

The models and observations (see references above) indicate that such a
transition occurs around an accretion rate $\mdot\sim0.01$.
Although the role of such radiatively inefficient states remains controversial, 
if such a change {\em does} occur, it would clearly rule 
out such a mode contributing to
the X-ray equivalents of bright quasars. As above, it would require
enormous BHs to produce such a luminosity at such low Eddington ratio
-- but RIAFs could easily contribute significantly to lower-luminosity
samples. A $\sim10^{8}\,\msun$ BH is still radiating at nearly
$10^{43}\,\ergsec$ in hard X-rays at $\mdot=0.01$, and the total
number density of such systems is $\sim10^{-2}\,{\rm Mpc^{-3}}$. Only
$\sim1-5\%$ of them need to be active at these low Eddington ratios to
account for the entire observed population of such X-ray sources in
the local universe \citep{ueda03:qlf}. But with rapidly declining
optical luminosity and disappearing broad line regions, such sources
may vanish from most Type 1 optical samples.

Together, these effects could lead to very different Seyfert
populations being observed in different bands or with different
selection criteria.  They also relate to a physical distinction in how
AGN may be fueled.
A low-mass BH at high Eddington ratio is likely to be in a gas-rich,
disk dominated system (given the typical properties of the 
expected low bulge-mass host). Fueling growth at these luminosities and masses
need not be related to violent events such as major mergers of
galaxies, but may be achievable by common mechanisms
such as the stochastic accretion of molecular clouds
\citep{hopkins:seyferts}, gas inflow in barred systems
\citep{jogee:review}, and minor mergers \citep{younger:minor.mergers}.
Although the fueling mechanisms can be relatively mundane, the
small-scale physics present are -- being systems at high Eddington
ratio -- simply a scaled-down version of quasar activity, and as such
represent a continuation of a feedback-regulated scenario, where
high-Eddington ratio activity (implying rapid BH growth) is truncated
by AGN feedback at a critical mass/luminosity corresponding to the
observed $M_{\rm BH}$-host spheroid correlations \citep{dimatteo:msigma}, 
and the AGN rapidly
decays to lower luminosities with some characteristic power law-like
light curve decline
\citep{hopkins:seyferts,hopkins:mdot.dist}.

On the other hand, a high-mass BH at a low Eddington ratio, living in
a massive host, is a natural consequence of a system already in its
late-time decay phase, having previously gone through a state of rapid
accretion, feedback, and self-regulated growth.  It is possible, of
course, to re-ignite such a system by the accretion of new cold gas in
a minor merger or via secular instabilities, but this will more likely
lead to a brief new burst of bright quasar activity before declining
again to a long-lived low Eddington ratio mode. Simulations and
observations suggest that systems can linger in this ``decaying'' mode
for $\sim$Gyr to a Hubble time \citep{yu:mdot.dist,
  volonteri:xray.counts,hopkins:lifetimes.interp,
  hopkins:lifetimes.obscuration,hopkins:groups.qso,
  yulu:lightcurve.constraints.from.bhmf.integration,
  ciotti:recycling.with.feedback.rad.vs.mech}, either in such a
post-merger steadily declining level of activity or eventually in an
equilibrium, accreting e.g.\ hot gas from the diffuse, pressure
supported gaseous halo in the spheroid or from stellar winds (the
distinction between this ``steady state'' accretion and a ``smooth
decline'' from initial high levels of activity triggered in e.g.\
mergers is largely semantic -- the two likely blend seamlessly into
one another).

Given the potential importance of these observational selection
effects, and the theoretical interest in understanding different modes
of accretion and their implications for observed AGN samples, it is of
considerable interest to develop predictions and observational
discriminants between these modes of accretion; it is also important
to recognize the observational selection effects that might lead to
the preferential selection of different populations.  Any observations
intending to study AGN host galaxies, fueling mechanisms, and
evolution must account for such distinctions.  In this paper, we adopt
a simple and general model for AGN lightcurves and use it to
illustrate the effects of dilution and radiative efficiency on
observational inferences of the Eddington ratio distributions,
obscured fractions, clustering, and host galaxy morphologies of AGN as
a function of luminosity and redshift.  Recognizing these biases,
however, the same observations, coupled with models such as those
presented here, can be used to break degeneracies between e.g.\
populations with different BH masses and Eddington ratios at fixed
luminosity; these can ultimately be used to constrain models for the
long-term evolution of AGN across many different stages of galaxy
evolution.

In \S~\ref{sec:mdot} we outline a simple semi-empirical way of
estimating the ``intrinsic'' BH mass/Eddington ratio distribution at a
given AGN luminosity; we further demonstrate how dilution and
radiative inefficiency can change this distribution. We also compare
to observations in different wavebands and discuss how the
observational inferences relate to theoretical models (taking into
account selection effects).  In \S~\ref{sec:obscuration} we examine
how dilution and radiative inefficiency influence estimates of the
``obscured'' or Type 2 AGN fraction.  In \S~\ref{sec:clustering} we
investigate how the observed clustering of AGN as a function of
wavelength, luminosity, and redshift can depend on the sample
selection criteria, and in \S~\ref{sec:morphology} we examine how
these effects can change the resulting host morphology distributions
at a given AGN luminosity. Finally, in \S~\ref{sec:discussion} we
summarize our results and outline observational tests that can resolve
the degeneracies highlighted in this paper.

For ease of comparison, we convert all observations to bolometric
luminosities given the appropriate bolometric corrections from
\citet{hopkins:bol.qlf} \citep[see also][]{richards:seds}, based on
the observations in \citet{elvis:atlas,
  george98:seyfert.line.spectra,vandenberk01:composite.qso.seds,
  perola02:compton.reflection.components,
  telfer02:mean.uv.qso.seds,ueda03:qlf,vbs03:alpha.ox}.  We adopt a
$\Omega_{\rm M}=0.3$, $\Omega_{\Lambda}=0.7$, $H_{0}=70\,{\rm
  km\,s^{-1}\,Mpc^{-1}}$ cosmology and a \citet{salpeter:imf} stellar
initial mass function (IMF), and normalize all observations and models
appropriately (note that this generally affects only the exact
normalization of quantities here, not the qualitative conclusions).
All magnitudes are in the Vega system.

\begin{figure*}
    \centering
    \plotside{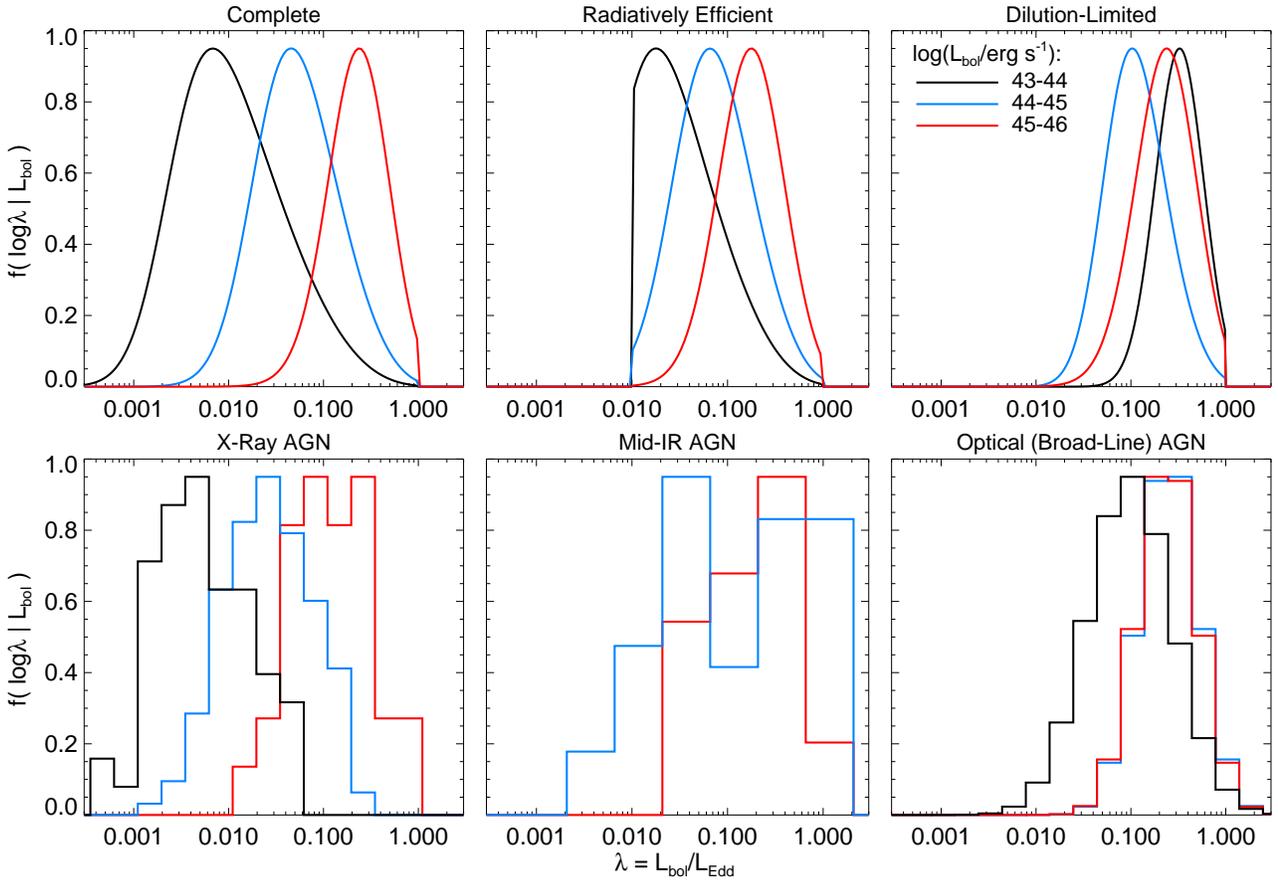}
    \caption{{\em Top:} Distribution of Eddington ratios (normalized
      ${\rm d}N/{\rm d}\log{\lambda}$, where $\lambda\equiv L_{\rm
        bol}/L_{\rm Edd}$) at a given AGN bolometric luminosity (as
      labeled), at $z\sim0-1$, derived from the models described in \S
      \ref{sec:mdot}.  
{\em Left:} ``True'' distribution -- i.e.\ that in a ``complete'' sample 
that can recover all AGN regardless of radiative efficiency and/or dilution 
(appropriate for deep X-ray and host-galaxy selected 
deep narrow-line surveys).  {\em Middle:} Distribution 
if only ``radiatively efficient'' AGN are detected, where 
we assume that thin disks disappear below $\dot m = 0.01$.  
This is appropriate for surveys relying on broad lines and/or
optical/IR continuum features (however deep).  {\em Right:}
Distribution if only AGN with optical/IR luminosities greater than
$\sim10\%$ of the host galaxy in the same band are selected. This is
required for any color/continuum selection in these wavelengths, but
also applies to most searches for broad-line signatures in complete
galaxy samples (especially if the width of those lines is
desired). Note the increasingly narrow range in the observed Eddington
ratio distribution -- and decreasing dependence on $L$ -- as these
selection effects are included. This is because low-$\lambda$ AGN are
potentially radiatively inefficient, and relatively less luminous
relative to their hosts.  {\em Bottom:} The observed Eddington ratio
distributions for observed samples of AGN at a given AGN bolometric
luminosity (based on bolometric correction from the X-ray or IR
luminosities). {\em Left:} X-Ray selected AGN from the observations in
\citet{hickox:multiwavelength.agn}, where the effects of radiative
efficiency and dilution are weak. This agrees well with the prediction
for a ``complete'' sample at each $L_{\rm bol}$. {\em Center:} Mid-IR
selected AGN, where radiatively inefficient AGN will be under-luminous
(and dilution effects enter as well, albeit to a lesser extent), also
from \citet{hickox:multiwavelength.agn}.  The distribution is less
luminosity-dependent, in reasonable agreement with predictions for
such a sample.  {\em Right:} Optically selected broad-line AGN from
\citet[][at $L_{\rm bol}>10^{44}\,{\rm erg\,s^{-1}}$]{kollmeier:mdot}
and \citet[][at $L_{\rm bol}<10^{44}\,{\rm
  erg\,s^{-1}}$]{greene:active.mf}; for these samples radiatively
inefficient AGN will not appear and AGN with luminosities much less
than their hosts will also be excluded.  Here there is almost no
dependence of Eddington ratio on luminosity, with all sources having
high $L_{\rm bol}/L_{\rm Edd}\sim0.1-1$, as predicted for samples with
similar observational selection criteria. \label{fig:ledd.dist}}
\end{figure*}

\section{Eddington Ratio Distributions and AGN Selection}
\label{sec:mdot}

In order to construct plausible estimates for the effects of dilution, 
we begin with a model for the AGN Eddington ratio distribution. 
We follow the methodology of \citet{hopkins:qso.all}: given the observed 
AGN luminosity function, a model for the average AGN 
lightcurve (or lifetime) uniquely translates to an Eddington ratio distribution 
as a function of luminosity and redshift. For example, if 
lightcurves were step-functions with some ``on'' and ``off'' $\mdot$, 
the Eddington ratio distribution would simply be the corresponding 
$\delta$-functions. Here, we adopt a simple 
parameterization of the lifetime/lightcurve motivated by 
hydrodynamic simulations with BH feedback 
\citep{hopkins:lifetimes.methods,hopkins:faint.slope}. 
The lightcurves, Eddington ratio distributions, and detailed comparison 
of both with observations are presented in \citet{hopkins:mdot.dist}; 
a number of other details and observational tests of the model (comparison 
with e.g.\ host mass distributions, clustering, and other properties) 
are presented in a series of papers \citep[e.g.][]{hopkins:lifetimes.interp,
hopkins:qso.all,hopkins:groups.qso}. 
We refer the interested reader to these works for more details, 
but briefly summarize the methodology here. 

We begin with the observed (redshift-dependent) bolometric 
quasar luminosity function, here from the compilation of observations in 
\citep{hopkins:bol.qlf} \citep[see also][and references therein]{shankar:bol.qlf}. 
We parameterize the AGN lifetime (the statistical representation of the 
lightcurve; namely, the differential time spent per unit luminosity, 
for a BH of a given mass) with a simple Schechter-function fit: 
${\rm d}t/{\rm d}\log{L} \approx
t_{0}\,(L/L_{0})^{-\alpha}\,\exp{(-L/L_{0})}$. 
The normalization
$t_{0}\sim 10^{8}$yr is the lifetime of high-Eddington ratio activity
and is comparable to the Salpeter time \citep{salpeter64} for the
$e$-folding of a BH at the Eddington limit; this estimate is also
reasonably consistent with observations (e.g., \citealt{martini04} and
references therein).
The ``cutoff'' at $L_{0}\approx0.4\,L_{\rm Edd}$ is the expected
cutoff near the Eddington rate for the given BH mass,
and $\alpha(M_{\rm BH})\sim0.5$ is a weak mass-dependent scaling
\citep[see][]{hopkins:mdot.dist}.

The formalism used here can represent a range of realistic AGN light
curve shapes, including those with an exponential rise/fall, those
with a range of variability power spectra, and those with a late-time
power-law like decay as predicted in a variety of self-regulated
growth scenarios \citep{menci:sam,granato:sam,
  sazonov04:qso.radiative.heating,hopkins:seyferts,ciottiostriker:recycling}.
Adjusting the model parameters to cover this broad range of
possibilities, we find that our qualitative conclusions are relatively
robust.  Given this, we use the values quoted above based on
hydrodynamic simulations.  Likewise, varying the choice of the bolometric 
QLF adopted by using just a subset of the observations compiled 
in \citet{hopkins:bol.qlf} makes little difference. 

Together, the bolometric luminosity function and 
prescription for AGN lifetimes 
specify the Eddington ratio distribution: the number at a 
given luminosity is known, and since the probability of a given system 
being at a given luminosity is trivially related to the shape of the lightcurve, 
one can de-convolve to determine the underlying BH mass/Eddington ratio 
distribution. Once this is specified, the most important aspect of the 
model is specified. Given the observed $M_{\rm BH}-M_{\rm bulge}$ 
relation \citep[with normalization and
scatter taken from][]{haringrix}, we can convert this to a host bulge mass
distribution. Likewise, given the observed correlation 
between galaxy mass and bulge-to-total mass ratio ($B/T$), 
and its scatter \citep[here from][]{balcells:bulge.scaling}, 
we can convert this to a distribution in total host galaxy mass 
and/or host morphology. Given some template galaxy SED 
or, more simply, an average stellar mass-to-light ratio as a 
function of galaxy mass (and redshift), well-determined 
in a number of studies \citep[see e.g.][]{bell:mfs,
conroy:hod.vs.z,ilbert:cosmos.morph.mfs,
vandokkum:tf.evolution, kassin:tf.evolution,
moster:stellar.vs.halo.mass.to.z1}, 
we trivially convert the host mass to a host luminosity distribution. 
Note that directly using the observed correlation 
between BH mass and total host luminosity \citep[e.g.][]{{kormendyrichstone95}} 
-- accounting for its larger scatter -- yields a nearly identical result. 

In \citet{hopkins:mdot.dist}, we compare the distributions of BH and
host properties inferred via the above procedure with the distribution
of Eddington ratios and BH and host masses measured in deep narrow
line samples at $z=0$ and inferred at $z=1$
\citep[from][]{heckman:local.mbh,hopkins:old.age,
  yu:mdot.dist,merloni:synthesis.model}.  The agreement is good over
the observed range from $M_{\rm BH}\sim10^{6}-10^{9}\,\msun$. 
Likewise, comparing with the host galaxy mass, luminosity, 
and color distributions in these low-redshift 
samples \citep{kauffmann:new.mdot.dist} and smaller high-redshift 
samples \citep[e.g.][]{peng:magorrian.evolution} yields reasonable agreement. So
these simple analytic models at least provide a good approximation to
the actual observed distributions in the local Universe. 

Given the above models, we take into account the effects of
dilution and/or radiatively inefficient accretion by considering three
simplified models for how the ``true'' distribution might be observed:
\begin{itemize}
\item {\bf (1)} Complete: All sources at the given bolometric
  luminosity are included, regardless of Eddington ratio or host
  properties. This might be representative of e.g.\ deep X-ray surveys 
  (modulo Compton-thick sources) 
  or narrow-line optical surveys.
\item {\bf (2)} Brighter than Host/Dilution-Limited: Given the
  observed dependence of galaxy mass-to-light ratios on stellar mass
  in various bands \citep{bell:mfs}, we convert our host bulge plus disk mass
  distribution to a host luminosity distribution as noted above, and compare it to
  the observed AGN luminosity in the same band \citep[with bolometric
  corrections from][]{hopkins:bol.qlf}. We include objects only above
  some threshold in the observed band; specifically, those
  sources for which the $B$-band luminosity of the AGN is larger than
  $10\%$ of its host luminosity in the same band.  This is appropriate
  for AGN samples selected on the basis of their broad-band IR-UV
  colors.
\item {\bf (3)} Radiatively Efficient only: We include sources with
  Eddington ratios above some critical threshold, $\mdot>\mdot_{\rm
    crit}$, where we take $\mdot_{\rm crit}=0.01$ (see
  \S~\ref{sec:intro}).  If there is indeed a change in accretion
  physics around this accretion rate, then such a cutoff is
  appropriate for samples that select AGN based on broad-band
  luminosities or colors in the UV, optical, or near infrared, or
  surveys that depend on identifying Type 1 signatures such as the
  presence of broad emission lines.
\end{itemize}

Figure~\ref{fig:ledd.dist} shows the inferred distribution of
Eddington ratios at various bolometric luminosities for the three
different sample cuts above.\footnote{In order to be consistent with
  typical observational procedures, the ``bolometric luminosity''
  plotted here is technically the bolometric luminosity that would be
  inferred based on the actual luminosity in some band 
  \citep[given the bolometric corrections from][used throughout]{hopkins:bol.qlf}. 
  If there is no
  qualitative change in SED shape with $\dot m$, this is the true
  bolometric luminosity, but in the ``radiatively efficient'' sample
  cut above, this represents instead the inferred luminosity assuming,
  as is typically done, a standard high $\dot m$ SED, even if the
  system is in fact radiatively inefficient and thus possesses an
  intrinsically different SED.}
At a given luminosity (unlike at a given BH mass), the Eddington ratio
distribution is roughly a lognormal distribution, albeit with some
non-negligible skewness \citep[as inferred by
e.g.][]{fine:broadline.distrib}. This is because, at sufficiently low
$\mdot$, arbitrarily high-$M_{\rm BH}$ BHs would be implied (at fixed
$L$), but the number of such systems is vanishingly small.
Figure~\ref{fig:ledd.dist} shows that the Eddington ratio distribution
can be dramatically affected by different observational limits,
becoming much narrower and more weakly dependent on luminosity when
luminosity/Eddington ratio cuts are applied.

\begin{figure}
    \centering
    \scaleup
    \plotter{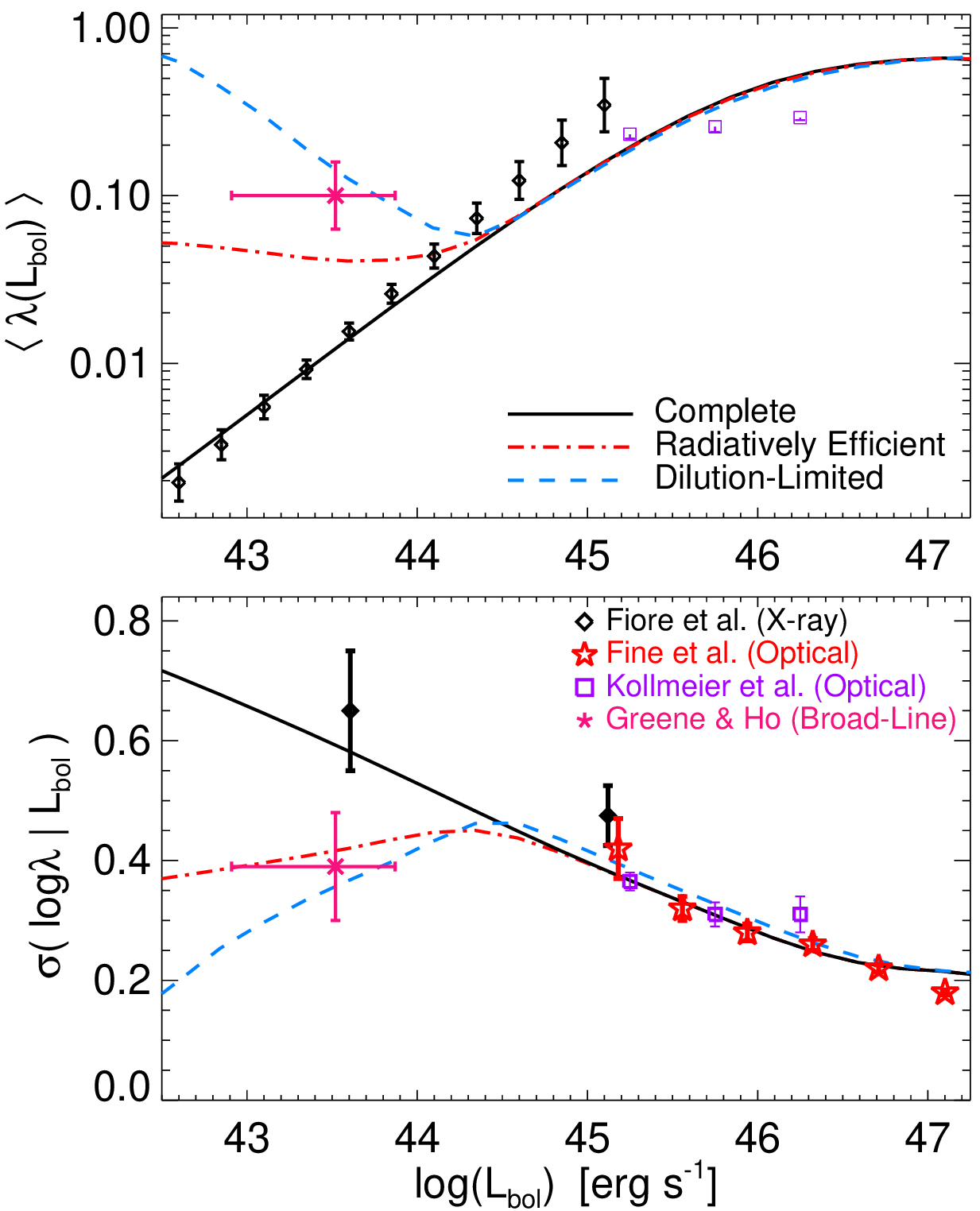}
    \caption{Median Eddington ratio $\lambda$ and $1\,\sigma$ dispersion 
    in Eddington ratio at a given bolometric AGN luminosity 
    (as in Figure~\ref{fig:ledd.dist}). 
    We compare the     predictions for the various selection criteria in Figure~\ref{fig:ledd.dist} 
    with the observed distributions inferred from the distribution of 
    X-ray to host luminosities in optically obscured AGN 
    \citep[][black diamonds]{fiore:type2.lx.vs.lhost,hasinger:absorption.update,
    hickox:multiwavelength.agn} (translating host luminosity to 
    average BH mass), and 
    from the distribution fitted to broad-line optical samples 
    in \citet{fine:broadline.distrib} from the 2dF, \citet{kollmeier:mdot} from AGES, 
    and \citet{greene:active.mf} using 
    the optical virial (line-width) BH mass estimators. 
    At high-$L$, all the selection criteria yield similar results in good agreement with the 
    observations -- almost all systems at these $L$ are radiatively efficient and bright 
    relative to their hosts. At low-$L$, the X-ray and broad-line 
    results disagree -- the low mean $\lambda$ and wide dispersion in 
    $\lambda$ near $L\sim10^{43.5}-10^{44}\,\ergsec$ 
    ($L_{2-10\,{\rm keV}}\sim0.3-1\times10^{43}\,\ergsec$) for the X-ray sample
    is consistent with the predictions for a complete sample, while the large $\lambda\sim0.1$ 
    and small dispersion in the broad-line sample of 
    \citet{greene:active.mf}  agrees with the predictions for a restricted 
    (dilution or efficiency-limited) sample. 
    \label{fig:ledd.dist.cum}}
\end{figure}

Figure~\ref{fig:ledd.dist.cum} shows the median and $1\,\sigma$
dispersions in Eddington ratio as a function of luminosity at $z =
0.5$, based on approximating the predicted distributions as lognormal
(we use the IPV width to prevent bias from outliers or skewness).
We compare the predictions with several observational estimates.

\citet{fine:broadline.distrib} consider the Type 1 quasar population
near $z\approx1$ in the 2dF survey, and estimate the distribution of
BH masses in narrow bins of luminosity relying on the commonly adopted
virial BH mass estimators 
\citep[see e.g.][and references
therein]{vestergaardpeterson:virial.corr.review}.  This allows them to
consider the distribution of BH masses via this proxy as a function of
luminosity down to near Seyfert luminosities.  Because the width of
the distribution can be determined without relying on the
(systematically still uncertain) absolute normalizations of these
calibrators, the authors decline to estimate absolute Eddington ratios
(although a rough estimate suggests they lie between $\sim0.1-1$, as
predicted here).  \citet{kollmeier:mdot} use the same technique over a
narrow luminosity range of Type 1 AGN in the AGES survey. At yet
lower-luminosities, \citet{greene:active.mf} study a sample of local
broad-line AGN from the SDSS down to extremely low luminosities
$L_{\rm bol}\sim10^{43}\,{\rm erg\,s^{-1}}$ ($L_{\rm
  H\alpha}\sim10^{6}\,L_{\sun}$). 

X-ray observations can study Type 2 objects at comparably low
luminosities, where the optical light is dominated by the host galaxy
and therefore a host galaxy stellar mass \citep[and corresponding BH
mass, adopting the observed $M_{\rm BH}-M_{\ast}$ relation
from][]{marconihunt} can be estimated. It is well-established that in
this regime the optical luminosity of the galaxy is approximately
constant while the X-ray luminosity changes, implying that at
lower-$L$ the X-ray luminosity function becomes largely a sequence in
Eddington ratio \citep{fiore:type2.lx.vs.lhost,
  barger:qlf.big,lafranca:hx.qlf,silverman:hx.selection,
  silverman:hx.spacedensity.ldde,silverman:hx.lf,
  hasinger:absorption.update,hickox:multiwavelength.agn}.

Figure~\ref{fig:ledd.dist.cum} shows the results of the X-ray sample
of \citet{fiore:type2.lx.vs.lhost}, where the optical $R$-band
luminosity is converted to a stellar mass based on the age and
mass-dependent observed mean $M/L$ ratios in
\citet{belldejong:disk.sfh} and \citet{bell:mfs}. We have
re-calculated these comparisons using the samples of
\citet{hasinger:absorption.update} and
\citet{hickox:multiwavelength.agn} and obtain the same result;
changing the assumed host $M/L$ within uncertainties makes little
difference.

Figure~\ref{fig:ledd.dist.cum} shows that the faint X-ray results
agree with optical results at high luminosities (as do the models).
At low luminosities, however, the X-ray results find that the median
accretion rate decreases and the scatter increases, in agreement with
our predictions for a ``complete'' sample.\footnote{We expect this 
to be independent of whether or not Compton-thick sources 
are missed in hard X-ray samples, since we are here assuming that 
whatever ``true'' obscuration is present is 
not host-property dependent. This may be uncertain, but if it 
were not so, it would only strengthen our conclusions.}
Narrow-line AGN samples also find a similar trend to that predicted 
for the ``complete'' sample. \citet{hao:local.lf.selection,hao:local.lf} find that 
host galaxy luminosity remains $\sim$constant 
as AGN luminosity declines, down to very low 
luminosities $L_{\rm bol}\sim 10^{42}\,{\rm erg\,s^{-1}}$. 
\citet{kauffmann:qso.hosts} make this explicit, with Eddington ratios 
distributions moving to lower $\mdot\lesssim 0.01$ and spanning 
an $\sim1\,$dex range at the lowest luminosities. 

However, the \citet{greene:active.mf} optical
results are quite different: they find a median Eddington ratio
$\mdot\sim0.1$ with a relatively small (factor $\sim2$) scatter,
similar to what is found in the optical broad-line samples at three
orders of magnitude higher luminosity. At the same bolometric
luminosity,
the median $\mdot$ is much higher than that in X-ray selected samples
and the spread is smaller. This result agrees well with the
expectation from our selection effect-limited models: because only
sources with clear optical broad-line signatures of a ``quasar'' are
selected, low-Eddington ratio systems which are 
simply underluminous relative to their hosts (and may 
also be radiatively inefficient) are excluded.

Figure~\ref{fig:ledd.dist} directly compares the Eddington ratio
distributions observed in different wavebands and the model
predictions, for AGN spanning three orders of magnitude in bolometric
luminosity. In the X-ray sample of \citet{hickox:multiwavelength.agn},
the observed distributions agree well with the expectation from the
``complete'' model -- in particular, the median Eddington ratio
clearly shifts to lower values with decreasing luminosity in the
manner predicted. On the other hand, in infrared or optically-selected
samples, the results are similar to that seen in
Figure~\ref{fig:ledd.dist.cum}: observed Eddington ratio distributions
are much narrower and less
luminosity-dependent.\footnote{\label{foot:ir.ledd}The observed
  infrared distributions are somewhat broader than the predictions;
  most of this can be attributed to scatter in the bolometric
  corrections and BH mass estimators. The predicted distributions
  would also be broader if the transition to radiative inefficiency
  were more gradual than we model.} This is more pronounced for the
optical sample than the infrared one, but this too could be
anticipated. In the optical, dilution is more severe than in the
infrared; however, radiative efficiency, 
if important, should influence the results in both
these bands similarly. 

\section{Implications for ``Obscured'' and Type 2 Populations}
\label{sec:obscuration}

Given the significant differences in the Eddington ratios selected in
different samples, we now study whether this will affect other
inferences drawn from AGN samples.  We first focus on the implications
for ``obscured'' and ``unobscured'' populations. Ideally, AGN would be
observationally classified as obscured on the basis of detailed
multiwavelength SED estimates -- in particular X-ray spectroscopy,
which can definitively show an absorption cutoff 
(although even in this case, Compton-thick sources will 
not be identified), and corresponding
absorption spectra in the infrared.  Although there are well-observed
systems where the presence or absence of actual dust/gas obscuration
can be unquestionably established \citep[e.g.][]{weymann:BALs,
  laor:warm.absorber, zakamska:multiwavelength.type.2.quasars}, these
observations are prohibitively time-consuming for the high redshift
surveys that are used to try to statistically quantify obscured
fractions as a function of e.g.\ luminosity and redshift. As a
consequence, AGN in these samples are typically classified as
``obscured'' on the basis of simplified proxies: the absence of
dominant AGN luminosity/colors (or a broad-line SED) in the optical,
or X-ray hardness. If however, the AGN emission is not dominant 
in the optical, these might
masquerade as ``obscured'' sources in observed samples.

\begin{figure}
    \centering
    \scaleup
    \plotter{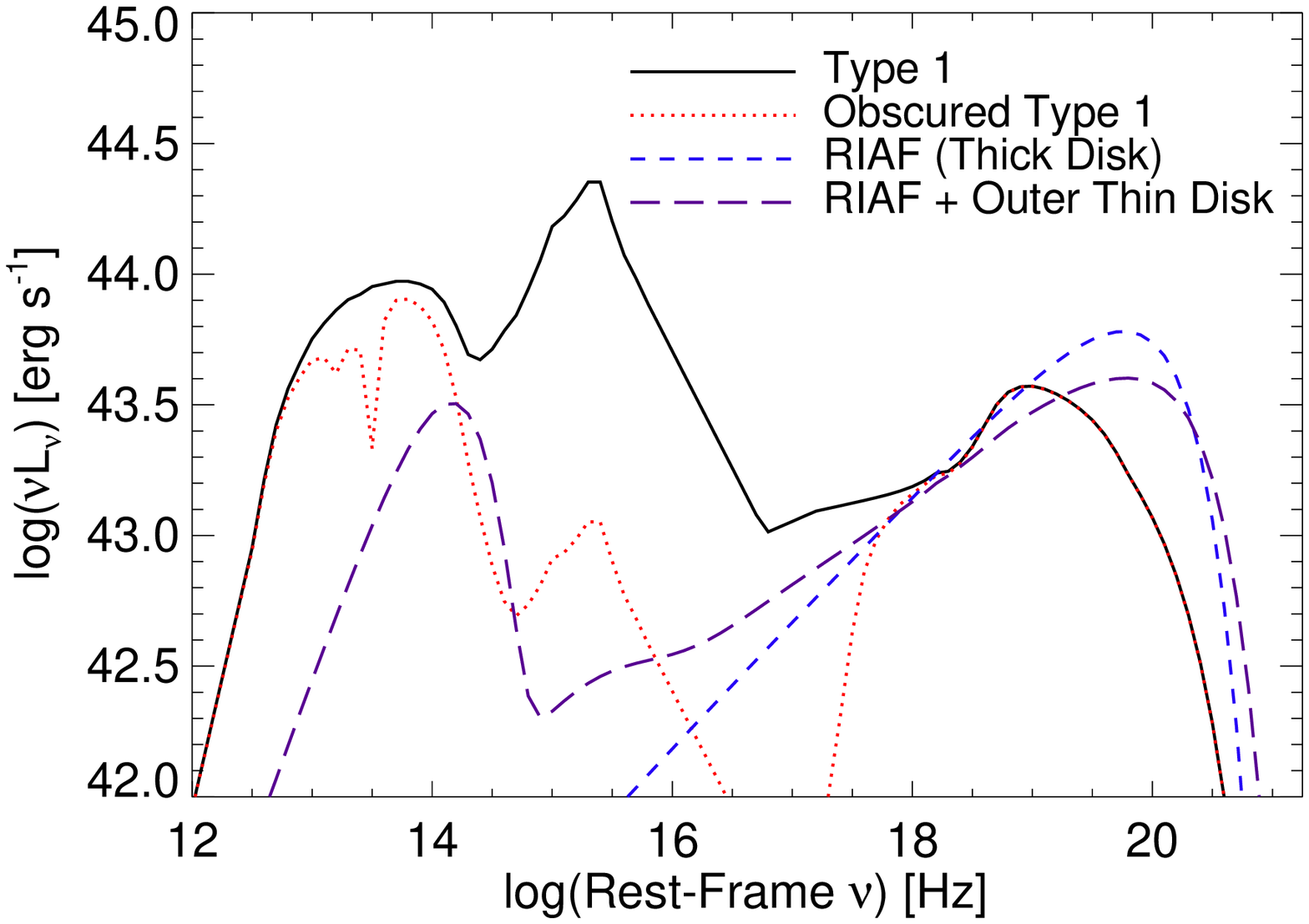}
    \caption{AGN SEDs: an unobscured, Type 1 QSO 
    \citep[from][]{hopkins:bol.qlf}, and that spectrum obscured by a column 
    with $N_{H}=10^{22}\,cm^{-2}$ \citep[with Milky Way gas-to-dust ratios and an 
    SMC-like reddening curve from][and a 
    small $\sim5\%$ optical/UV scattering component]{pei92:reddening.curves}; 
    we also show 
    the  predicted spectrum of an optically thin RIAF 
     \citep{quataert:adaf.spectral.models} and a model with 
    an inner RIAF and outer (truncated) thin $\alpha$-disk.
    Without detailed SEDs, the RIAF models are indistinguishable from 
    obscuration in the optical, and yield 
    power-law like X-ray spectra with hardnesses similar to 
    obscured Type 1 objects. The 
    luminosities are arbitrarily scaled to be comparable 
    to the systems of interest in X-rays.
    \label{fig:compare.templates}}
\end{figure}

Figure~\ref{fig:compare.templates} illustrates these issues. We first
compare template SEDs of a Type 1 quasar and a ``true'' obscured
quasar.  The obscured system is the product of the Type 1 template
attenuated by a column of $N_{H}=10^{22}\,{\rm cm^{-2}}$, with a Milky
Way-like gas-to-dust ratio, solar metallicity, SMC-like dust reddening
curves fitted in \citet{pei92:reddening.curves}, and photoelectric
absorption in the X-ray following
\citet{morrison.mccammon.83:photoelectric.absorption}; for details see
\citet{hopkins:qso.all}.
Differences in e.g.\ the gas-to-dust ratios and reddening curves of
AGN will somewhat alter the quantitative details here, but our
qualitative conclusions below are not sensitive to these choices.

Dilution can easily hide an AGN in the optical or infrared; recall
(eq.~\ref{eqn:dilution}) that at an Eddington ratio $\mdot\sim0.01$,
even if the system is still radiatively efficient, the $B$-band
luminosity of the AGN is $<10\%$ of the host galaxy luminosity in a
relatively old elliptical host galaxy (with high $M_{\ast}/L$ and
$B/T$); in a young spiral galaxy \citep[with typical
$B/T\lesssim0.2$;][]{weinzirl:b.t.dist} that might host a
$\sim10^{6}-10^{7}\,\msun$ BH, the fraction of the optical light drops
to $<1\%$.  At these ratios, the galaxy continuum dominates, and the
system would be classified as Type 2 or obscured according to any
color, continuum, or broad-band based photometric criteria. 
The ``true'' AGN SED is unobscured, but the optical/IR light is 
predominantly galaxy light (by definition, in this regime). Broad
lines would in principle be detectable, but only with very deep
observations -- the peak broad line height in the optical SED, given
the composite quasar SED \citep[see][]{vandenberk01:composite.qso.seds},
would appear as a $<2\%$ correction to the continuum for the
star-forming case above \citep[comparing to model SEDs from][]{BC03}.
This is prohibitively time-consuming for all but a few local,
well-studied systems \citep[even in the early type galaxy, reliable
identification of a single broad line would require $S/N\sim 30$
spectroscopy; typical SDSS spectra, for comparison, are $S/N\sim6$,
leading to difficulty making broad-line identifications when the AGN
fraction falls below
$\sim30-50\%$;][]{vandenberk:qso.spectral.decomposition}.  In
practice, most objects in the large samples used to classify systems
as obscured or unobscured require that the AGN continuum dominate in
the optical in order to define a system as Type 1 (in detail, systems
are often classified as Type 1 or 2 on the basis of whether a single
galaxy or AGN template provides a better fit to the optical; in such a
case, the requirement that the AGN dominate the light is explicit).

Figure \ref{fig:compare.templates} further compares the Type 1 and
true obscured SEDs with two template spectra for radiatively
inefficient accretion models.  First, we consider a RIAF around a $5
\times 10^{8}\,\msun$ BH with a dimensionless accretion rate of $\mdot
= 0.01$.  The dimensionless spectrum $\nu L_{\nu}/L_{\rm Edd}$ is a
weak function of black hole mass at fixed $\mdot$, and so can be
trivially scaled to different luminosities (in
Figure~\ref{fig:compare.templates} we arbitrarily scale the SEDs to be
comparable in X-ray luminosity, since we will be considering samples
observed in fixed luminosity bins).  The spectral calculations here
are based on the models of \citet{quataert:adaf.spectral.models}, to
which we refer the reader for details.  The one important difference
is that we use the results of \citet{sharma:electron.heating.in.adafs}
for electron heating in RIAFs.  One uncertainty in the modeling is
that RIAFs have strong outflows that may carry away a significant
fraction of the mass supplied at large radii \citep{blandford:adios};
for the models of interest here, the emission is largely produced at
small radii close to the black hole horizon, and so the spectrum is
not strongly sensitive to the presence of outflows so long as the
accretion rates quoted here are interpreted as $\mdot$ at small radii
close to the black hole's horizon; equivalently, we can consider
systems at fixed luminosity regardless of the large-scale accretion
rates producing that luminosity. In short, this affects only the ratio
of the mass accretion rate to luminosity; since we will consider
objects at fixed luminosity, this makes no difference.

At the accretion rates relevant to this paper, theoretical models and observations 
of both X-ray binaries \citep[e.g.][]{esin:truncated.thin.disk.xrb} and 
AGN \citep[e.g.][]{quataert:adaf.thin.disk.truncation} 
suggest that the inner RIAF is likely surrounded by an outer thin disk. 
The latter produces a non-negligible optical-IR flux, although much 
less than a standard radiatively efficient disk would produce if it extended 
close to the black hole. In addition, the disk photons inverse Compton 
cool the inner RIAF, softening the X-ray spectrum.
This is shown explicitly in Figure~\ref{fig:compare.templates} 
where we also plot the spectrum for a model in 
which the transition from thin disk \citep[modeled as a 
standard][$\alpha$-disk]{shakurasunyaev73} to RIAF occurs 
at 100 Schwarzschild radii.

In either case, it is straightforward to see how the RIAF spectra
would be observationally equivalent to an obscured AGN in the
optical. Interestingly (as we discuss further below), the RIAF spectra
are also hard power-laws in the X-rays: their hardness can be similar
to a gas-obscured intrinsic Type 1 AGN spectrum.

\begin{figure*}
    \centering
    \scaleup
    \plotside{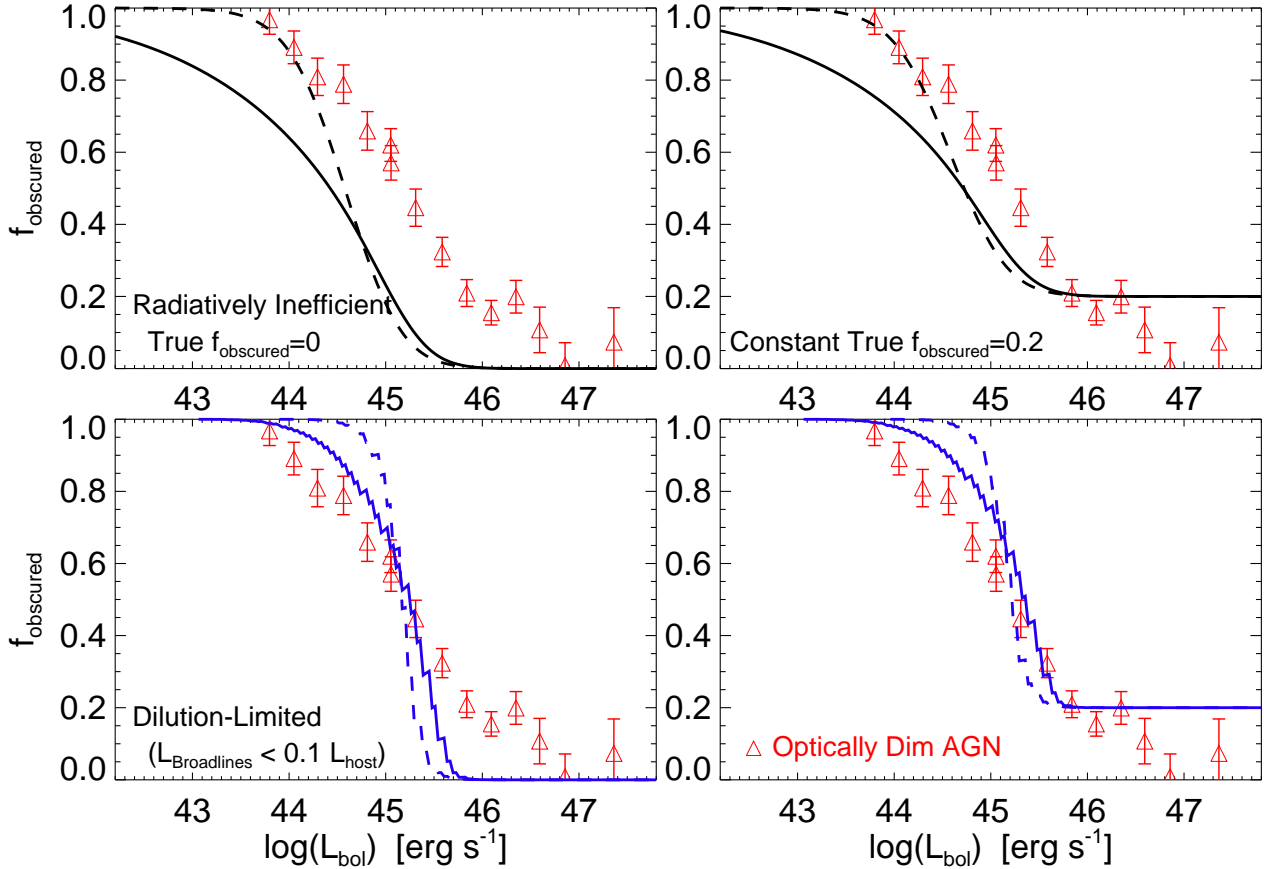}
    \caption{Predicted fraction of AGN at a given bolometric 
    luminosity that would be classified as ``obscured'' 
    in optical samples. {\em Top Left:} Fraction of objects at a given bolometric 
    luminosity that are radiatively inefficient and thus 
    would be absent in optical samples. Solid and dashed lines correspond to the allowed 
    range of Eddington ratio distributions 
    derived from models and observations \citep[see][]{hopkins:mdot.dist}. 
    We compare with the observed 
    ``obscured'' fraction from the 
    observations in \citet{hasinger:absorption.update}, 
    where (X-ray selected) objects are labeled ``obscured'' based on the lack of 
    dominant broad-line AGN  in the optical. 
    {\em Top Right:} Same prediction, but with a constant 
    (luminosity-independent) 
    fraction $=0.2$ of genuinely obscured sources added at all $L$. The
    combination of a true 20 \% obscured fraction and radiatively 
    inefficient accretion at low $L$ can approximately 
    reproduce the observationally inferred ``obscured'' sources. 
    {\em Bottom Left:} As top left, but showing the 
    predicted fraction of systems which will drop out in the optical from a dilution-limited 
    sample (or samples where Type 1/2 classification is based on optical/UV/IR colors, 
    continuum, presence of broad-lines, or 
     goodness-of-fit of template AGN versus galaxy SEDs in the optical/IR). 
    {\em Bottom Right:} As bottom left, with a constant fraction $=0.2$ of 
    genuine obscured sources.  The observed trend of ``obscured'' fractions 
    with luminosity can be explained  primarily by these Eddington 
    ratio effects, with a true obscured fraction $\sim 0.2$ is that is 
    much less dependent on luminosity than a direct interpretation of the data would imply. 
    \label{fig:obscuration}}
\end{figure*}

\begin{figure}
    \centering
    \scaleup
    \plotter{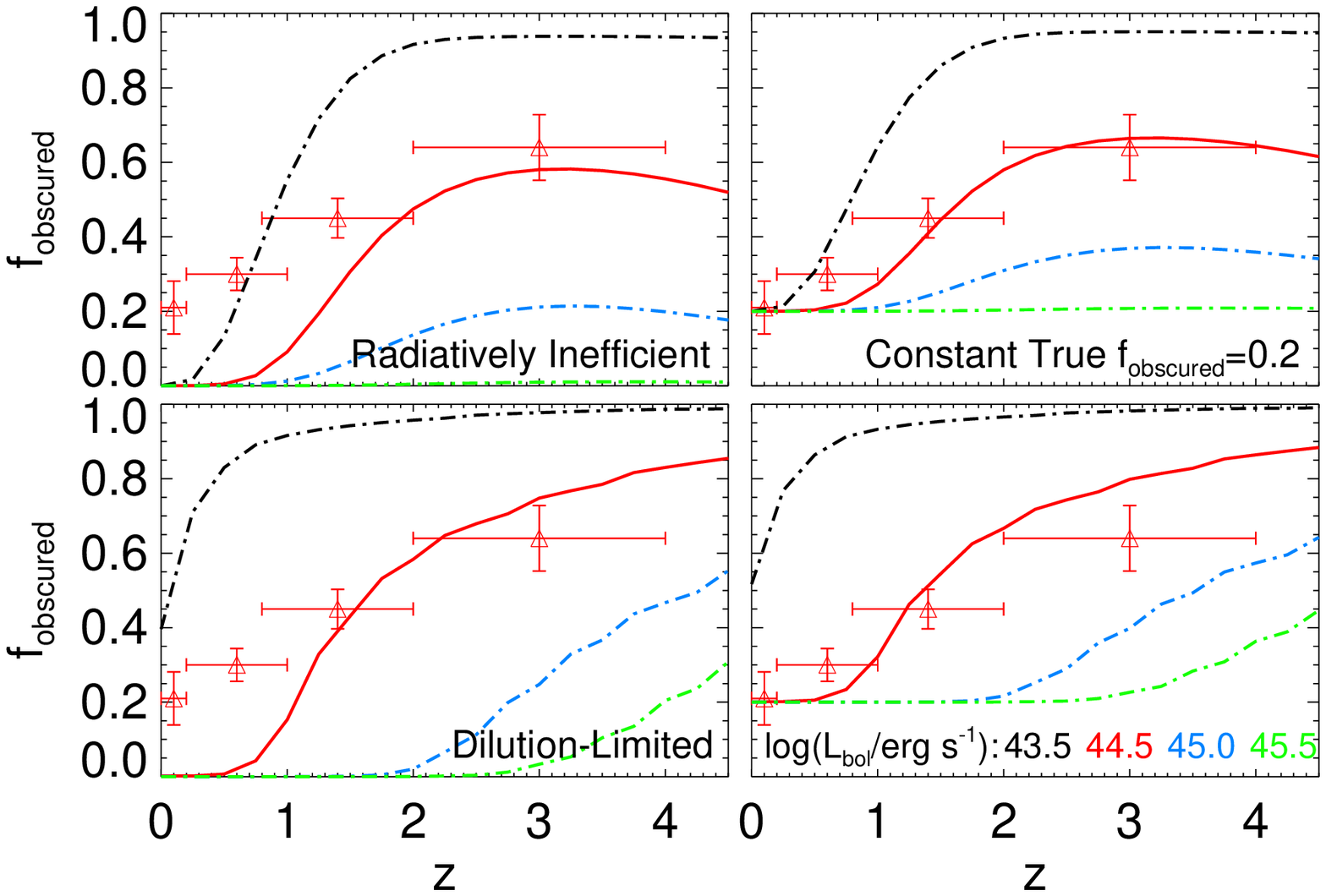}
    \caption{As in Figure~\ref{fig:obscuration}, but plotting the 
    predicted fraction  that would appear ``obscured'' in optical 
    samples at a given  $L$ (as labeled)
    as a function of redshift. Also shown are the corresponding observational
    estimates from \citet{hasinger:absorption.update} for 
    luminosities in the range $L\sim10^{44}-10^{45}\,\ergsec$. 
     \label{fig:obscuration.z}}
\end{figure}

Given our models for the Eddington ratio distributions of AGN, we can
estimate the fraction of systems that would erroneously be classified
as optically ``obscured'' because of either dilution or, more speculatively, 
a transition to radiatively inefficient accretion. This is simply the 
fraction, relative to a ``complete'' sample, that meet our 
previous criteria {\bf (2)} or {\bf (3)}, respectively. 
Figure~\ref{fig:obscuration} shows the resulting ``obscured'' fraction
as a function of luminosity and Figure~\ref{fig:obscuration.z} shows
the ``obscured'' fraction as a function of redshift.  Because at lower
luminosities the fraction of low-Eddington ratio systems is larger
(see \S~\ref{sec:intro}), this generically leads to a decreasing
``obscured'' fraction with increasing luminosity.  Moreover, because
the shape and break location of the luminosity function evolve with
redshift 
\citep[implying that the distribution of Eddington ratios 
shifts accordingly in high-mass BHs; recall here we adopt 
the QLF evolution from the compilation in][but 
other observations yield similar results]{hopkins:bol.qlf}, 
at fixed luminosity there can be a weaker but still non-zero
dependence of the ``obscured'' fraction on redshift
(Fig. ~\ref{fig:obscuration.z}). Specifically, 
it is well-established that more massive BHs are preferentially 
active at higher redshifts; as such, it is possible that a population at 
fixed (sub-$L_{\ast}$) luminosity includes a larger fraction of high-mass BHs at 
low Eddington ratio, leading to larger (fractional) dilution. In addition, 
for the same host galaxy mass, star formation rates (and therefore stellar 
population luminosities) increase with redshift. 

The trends we find are similar to the increasingly well-established
observed trends in various samples \citep{hill96:sf.abs.in.radio.gal,
  simpson99:thermal.imaging.of.radio.gal,
  willott00:optical.qso.frac.vs.l,simpson00:ir.photometry.radio.qsos,
  steffen03:agn.obsc.vs.l.z,ueda03:qlf,grimes04:obsc.frac.vs.l,hasinger04:obsc.frac.and.xrb,
  sazonov.revnivtsev04:local.hx.qlf,barger:qlf,
  simpson:type1.frac,hao:local.lf,gilli:obscured.fractions,
  hickox:bootes.obscured.agn,hasinger:absorption.update}.  We compare
  the observations from \citet{hasinger:absorption.update}, 
  a hard X-ray selected sample, to our simple estimates in
Figures~\ref{fig:obscuration} \&\ \ref{fig:obscuration.z}.  Although
some of the observationally identified ``obscured'' systems are almost
certainly obscured in the real, traditional sense (especially at the
highest luminosities), our estimates suggest that a reasonable
fraction of the very high inferred obscured fractions (approaching
unity) may owe to these selection effects, rather than to actual
dust/gas obscuration. 

In detail, even with the strong assumptions in
Figures~\ref{fig:obscuration} \&\ \ref{fig:obscuration.z}, that all
objects below $L_{B,\ {\rm AGN}}/L_{B,\ {\rm host}}<0.1$ (dilution-limited) 
or below $\mdot=0.01$ (radiative efficiency-limited) 
are classified as ``obscured,'' we do not reproduce
the full observationally inferred obscured fraction.  
Including a constant (luminosity and redshift-independent) ``true''
obscured fraction of $0.2$, our inferred ``obscured'' population is
similar to observational estimates as a function of both luminosity
and redshift (right-hand panels in Figs. \ref{fig:obscuration} \&
\ref{fig:obscuration.z}). Thus, although some genuine obscuration is
required, it might be as low as $\sim 20\%$ of the population,
significantly less than what is typically inferred in low-luminosity
AGN samples; more important, the majority of the inferred luminosity and
redshift dependence in observed samples can be accounted for by the
effects of a non-trivial Eddington ratio distribution and the
inevitable selection effects in optical continuum or broad-line
samples. 

Note that we are not arguing that the obscured fraction has to be 
as low as $\sim20\%$. Given the uncertainties (error bars in the 
observations, range of allowed Eddington ratio distributions and systematic 
uncertainty in our simple model of host properties), it could be somewhat 
(factor $\sim2$) higher \citep[as suggested by other 
independent constraints from 
e.g.\ infrared emission or 
X-ray background fitting; see e.g.][]{treister:obsc.frac.from.midir.excess}.
What we show, however, is that simple effects 
could, within reasonable parameter space, account for up to 
$\sim80\%$ of the claimed effects, in particular, 
the strong dependence on luminosity and redshift. 
In fact, direct observations that do attempt 
to more rigorously distinguish obscured and unobscured sources 
in ``complete'' samples (e.g.\ narrow-line studies) favor a 
somewhat higher ``true'' obscured fraction $\sim30-40\%$, 
but with a similar lack of luminosity and/or redshift dependence 
\citep[see e.g.][and references therein]{hao:local.lf.selection,hao:local.lf}. 

\begin{figure}
    \centering
    \scaleup
    \plotter{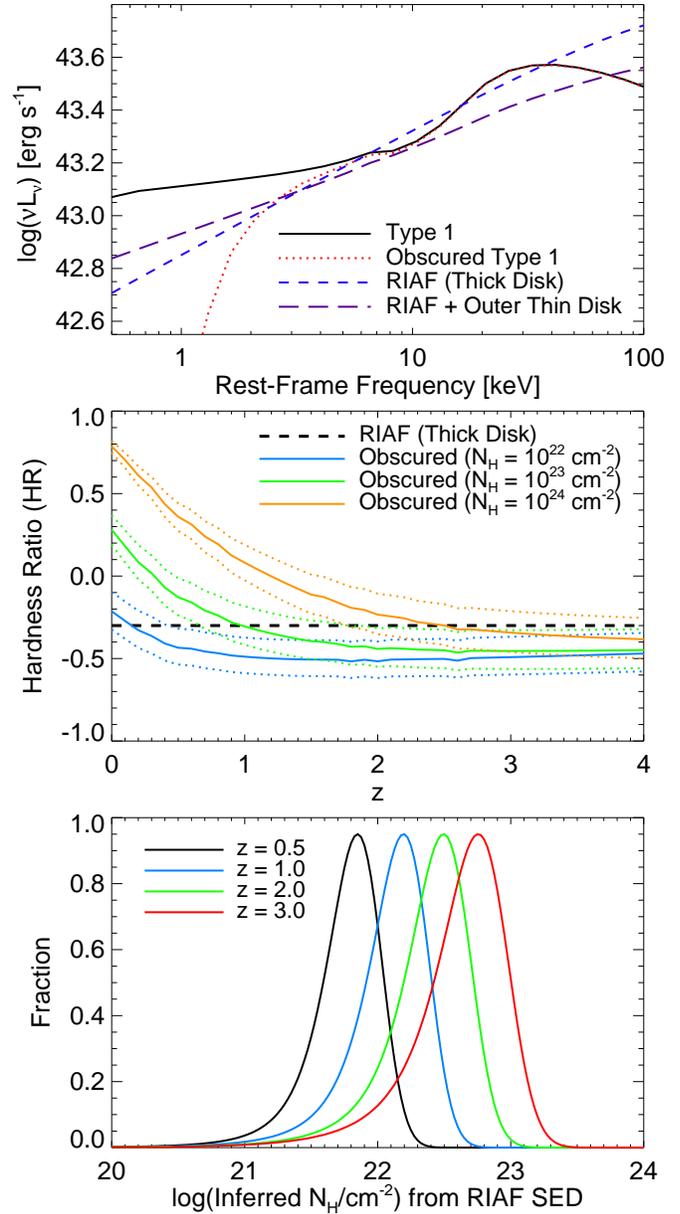}
    \caption{{\em Top:} Comparison of template AGN SEDs in the X-ray;
      we conside Type 1, obscured, pure RIAF, and RIAF+outer disk
      models.  The RIAFs predict hard X-ray SEDs similar to an
      obscured system, but without a true absorption cutoff. {\em
        Middle:} Observed hardness ratio (HR) from the templates in
      the top panel as a function of redshift. The Type 1 template is
      obscured by different columns
      $N_{H}=10^{22},\,10^{23},\,10^{24}\,{\rm erg\,s^{-1}}$ for input
      photon index $\Gamma=1.9$ (solid lines) and $\Gamma$ varying
      from $1.8-2.0$ (dotted lines). At $z\gtrsim1$, the strong
      absorption cutoff in true obscured systems is redshifted out of
      the soft band, and the hardness ratios from obscured systems are
      similar to the power-law spectra expected in RIAF models.  {\em
        Bottom:} Inverting the {\em middle} plot: Given the template
      RIAF hardness, we plot the column density $N_{H}$ that would be
      inferred by matching this hardness to an obscured Type-1 SED.
      We assume observational errors of $\sim0.1$ in HR, comparable to
      deep observed X-ray samples
      \citep[see][]{hasinger:absorption.update}.
      \label{fig:hardness.of.templates}}
\end{figure}

Qualitatively similar observational results 
to those shown in Figure~\ref{fig:obscuration} on the fraction of obscured systems have
been obtained by using X-ray hardness (rather than optical SEDs) as a
proxy for obscuration; i.e, systems with sufficiently hard X-ray
spectra are classified as obscured (although it should be noted, 
the statistics in these studies are much more limited). 
If real, this is difficult to explain with simple host galaxy dilution. 
However, in Figure~\ref{fig:compare.templates}, we noted
that the expected RIAF spectra are similar in their X-ray hardness to
a genuine obscured (intrinsically Type 1) SED. We consider this in
more detail in the top panel of
Figure~\ref{fig:hardness.of.templates}, which shows the X-ray SEDs for
Type 1, genuinely Type 2 dust/gas-obscured, pure RIAF, and RIAF+outer
thin disk models.The RIAF SEDs are hard power laws; they do not 
have an absorption cutoff. But, if we ignore that cutoff, only seen 
at $<1\,$keV rest-frame, the RIAF SEDs are sufficiently hard that they are 
indistinguishable (with typical signal-to-noise) 
from the SED of the true obscured AGN; both are 
significantly harder than the SED of an unobscured AGN out to $>10\,$keV. 

To quantify the extent to which our theoretically calculated RIAF
spectra could mimic obscuration,
Figure~\ref{fig:hardness.of.templates} also shows the hardness ratio
(HR) as a function of galaxy redshift for these template SEDs, with
hardness ratio defined in terms of the observed-frame hard (2-10 keV)
and soft (0.5-2 keV) band counts, HR$=(H-S)/(H+S)$.\footnote{Note that
  the observed hardness ratio, as defined in terms of counts, depends
  on the exact waveband and instrument response function. For the sake
  of generality, what we quote here can be thought of as an
  ``intrinsic'' HR for an instrument with perfect response and
  sensitivity, with fixed bands defined from $0.5-2.0$\,keV and
  $2.0-10.0$\,keV; as such, it is not directly comparable to hardness
  ratios in the literature for real instruments and wavebands.
  Because we are interested in comparing two input SEDs (the obscured
  AGN and RIAF), however, so long as we treat them identically it
  makes little difference to our conclusions whether we adopt this
  definition or model a more complex instrumental response function.
  There is, however, some weak sensitivity to the wavebands in which
  hardness ratios are defined, since these will sample various
  portions of the absorption cutoff at different redshifts.}  Larger
absolute values of HR correspond to ``harder'' SEDs and, it is usually
believed, a more obscured system.

The RIAF model, being a near power-law, predicts a constant,
relatively hard HR. We compare this to the HR derived from the Type 1
template shown, obscured by different columns
$N_{H}=10^{22},\,10^{23},\,10^{24}\,{\rm erg\,s^{-1}}$. The template
shown corresponds to an intrinsic photon index $\Gamma=1.9$
\citep[see][]{elvis:atlas,george98:seyfert.line.spectra,
perola02:compton.reflection.components,ueda03:qlf,vbs03:alpha.ox},
with a reflection component generated by the PEXRAV model
\citep{magdziarz.zdziarski.95:compton.reflection.model} following
\citet{ueda03:qlf}. We also show the results for varying $\Gamma$,
within the range suggested by observations, $\Gamma=1.8-2.0$ (dotted
lines in the middle panel of Fig. \ref{fig:hardness.of.templates}).

At $z\gtrsim1$, the strong absorption cutoff in true obscured systems
is redshifted out of the soft band, and the hardness ratios from
obscured systems are similar to the power-law spectra expected in RIAF
models (middle panel; Fig.~\ref{fig:hardness.of.templates}).  We
therefore consider what column densities would be inferred, if the
true spectrum were the RIAF model observed at some redshift, but the
observed hardness ratio were modeled as the template Type 1 SED plus
obscuration (allowing for typical errors in hardness ratio $\Delta
HR\sim0.1$, typical of the quality in deep X-ray samples). This is the
common method for inferring obscuration in X-ray samples
\citep[e.g.][]{ueda03:qlf,lafranca:hx.qlf,
  barger:qlf,gilli:obscured.fractions}; applied to a hard power-law
like spectrum expected in radiatively inefficient states, it can lead
to a significant false obscuration signal, with inferred column
densities $N_{H}\sim 10^{22}-10^{23}\,{\rm cm^{-2}}$ (bottom panel of
Fig.~\ref{fig:hardness.of.templates}). At these redshifts, observations 
are sampling the regime where the RIAF SEDs are significantly harder than 
that of an unobscured AGN, but missing the regime where the 
strong absorption cutoff distinguishing true absorption can be sampled. 

\begin{figure}
    \centering
    \scaleup
    \plotter{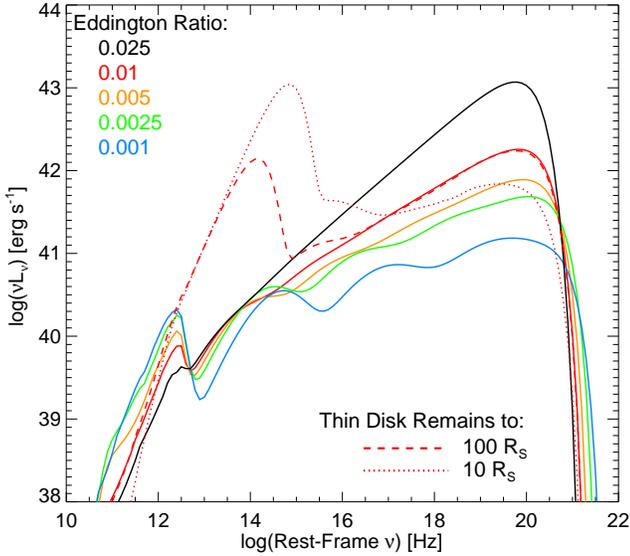}
    \caption{Model RIAF spectra at Eddington ratios 
    $\mdot=0.025,\,0.01,\,0.005,\,0.0025,\,0.001$ (solid; black, red, orange, 
    green, blue, respectively, from top to bottom in 
    the X-rays). Also shown are models with $\mdot=0.01$ in which 
    there is an outer thin disk that transitions to the inner RIAF 
    at a radius of 10 (dotted) and 100 (dashed) Schwarzschild radii.
    For the simple comparisons made in this paper 
    (``obscured'' or ``unobscured''), the differences between 
    the different $\dot m$ have little impact; the optical 
    SEDs are always strongly suppressed, and the X-ray 
    SEDs are power-law like (over $\sim0.5-100\,$keV) with similar hardness 
    for the accretion rates of interest, down to very low-$L$ 
    ($L_{\rm bol}\ll 10^{42}\,{\rm erg\,s^{-1}}$) where 
    the spectra become softer.
    \label{fig:riaf.templates}}
\end{figure}

Figure~\ref{fig:riaf.templates} shows how the RIAF spectrum varies as
a function of accretion rate. As $\mdot$ decreases, the predicted
X-ray spectra become softer \citep[a transition that has also been
suggested from observed AGN; see
e.g.][]{dai:xr.hardness.vs.mdot,saez:xr.hardness.vs.mdot,
  constantin:xrhard.vs.mdot}; however, they are still harder than an
unobscured AGN for all $\mdot\gtrsim0.001$ (below which the
luminosities are sufficiently low that they contribute negligibly to
the different fixed-luminosity samples considered here). It therefore
makes little difference whether we simply pick one of these Eddington
ratios as ``representative'' of the RIAF SED, or include the full
$\mdot$ dependence in the predictions here.  Changing other parameters
in the model shown in Figure~\ref{fig:riaf.templates}, such as the
disk viscosity, the magnetic energy density (relevant for the
predicted synchrotron emission), and mass loading of outflows does not
significantly change the spectral shape at fixed $\mdot$.  The
strength of outflows is important for the normalization of the SED,
i.e., for the ratio of the luminosity to the (large-scale) accretion
rate, but at fixed luminosity makes little difference. In other words,
although some of the details of RIAF accretion flows remain uncertain,
the vast majority of these issues do not affect our qualitative
conclusions. RIAF SEDs are generally X-ray hard; this must be true,
since they are designed to describe the observed ``low/hard''
(low-luminosity, hard X-ray spectrum) state in X-ray binaries.  The
dominant uncertainty for our purposes lies in the nature of the
transition to the RIAF state. If this is more gradual than what we
have assumed here -- i.e.\ if a significant thin disk survives all the
way down to small radii for Eddington ratios $\mdot<0.01$, then the
X-ray SED could be significantly softer (owing to inverse Compton
cooling of the inner thick disk photons), although it must survive to
very small radii indeed (only being an important effect if the
transition radius is $\lesssim 10$ Schwarzschild radii).  Of course,
this is essentially equivalent to the uncertainty regarding whether
such a transition occurs at all \citep[see, for
example,][]{maoz:liners.still.have.normal.uv.properties}.

It is also important to note, as is clear in Figure~\ref{fig:hardness.of.templates} 
that the RIAF SEDs do not become extremely hard. It is difficult, with such 
models and no true obscuration, to mimic 
e.g.\ Compton thick columns with $N_{H}>10^{23}\,{\rm cm^{-2}}$. Ultimately, 
more detailed observational studies will allow direct comparison of the 
hardness ratio distribution at each redshift, which could include components 
that cannot be mimicked by RIAF models. 

\begin{figure*}
    \centering
    \scaleup
    \plotside{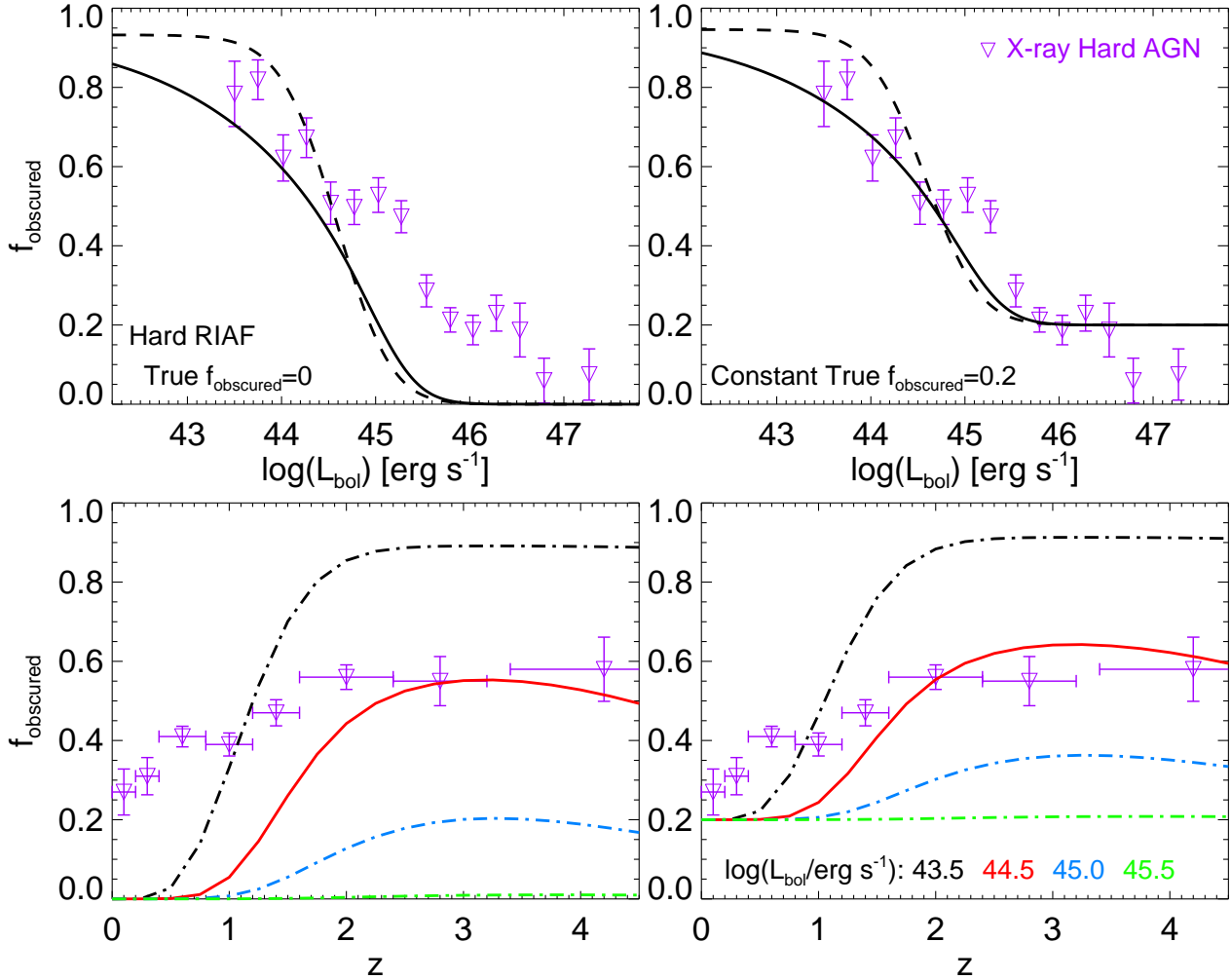}
    \caption{The fraction of systems that would be classified as ``obscured'' based on their
    observed X-ray hardness ratios (as shown in 
    Figure~\ref{fig:hardness.of.templates} as a function of luminosity 
    and redshift); we assume that for $\mdot<0.01$, all 
    accretion is via a RIAF.  The predictions are compared with the observed fraction of hard 
    X-ray  sources, typically classified as ``obscured''  
    \citep{hasinger:absorption.update}.  Because RIAFs 
    have intrinsically hard spectra, X-ray hardness alone 
    does not necessarily break the degeneracies between 
    selection effects and true obscuration demonstrated 
    in Figures~\ref{fig:obscuration}-\ref{fig:obscuration.z}.  
    X-ray spectra with the sensitivity to  measure the presence or absence of 
    a true absorption cutoff are needed to definitively show obscuration. 
    \label{fig:obscuration.hardness}}
\end{figure*}

We now estimate the fraction of AGN that might be classified as
``obscured'' on the basis of observed X-ray hardness, as a consequence
of having an intrinsic RIAF-type SED. This is shown in
Figure~\ref{fig:obscuration.hardness} as a function of luminosity and
redshift.  The model is the same as in our predictions for optical
samples, but we assume that all sources with $\mdot<0.01$ have a thick
disk spectrum as in Figure~\ref{fig:compare.templates}, and calculate
the hardness ratio distribution and implied column density
distribution as in Figure~\ref{fig:hardness.of.templates}.  As in the
observational literature (e.g., \citealt{gilli:obscured.fractions} and
references therein), we classify as ``obscured'' all sources with an
inferred column $N_{H}>10^{22}\,{\rm erg\,s^{-1}}$. The radiatively
efficient ($\mdot>0.01$) sources are assumed to be either uniformly
unobscured (``true $f_{\rm obscured}=0$'' case; left panel)
or to have a constant fraction $f_{\rm obscured}=0.2$ of sources
genuinely obscured (right panel).  The resulting trends in obscured
fraction as a function of luminosity and redshift are similar to those
in Figures~\ref{fig:obscuration} \&\ \ref{fig:obscuration.z} (the
optical samples). By taking into account the predicted change in AGN
SEDs below $\dot m \simeq 0.01$, we thus conclude that a significant
fraction of the inferred dependence of obscuration on
luminosity/redshift, both determined from optical observations and
X-ray hardness cuts, could owe to the fraction of systems in a
radiatively inefficient state, rather than from true obscuration.

\section{AGN Clustering versus Selection and Fueling}
\label{sec:clustering}

By affecting the distribution of Eddington ratios detected in a given
sample, dilution and radiative efficiency selection effects can also
affect the BH masses to which the sample is sensitive, and thus
stellar/host masses.
Given the strong dependence of clustering on stellar/halo mass, the overall
clustering amplitude measured in such a sample can be different than
that inferred from a sample including all active systems.

Another important effect for observed clustering is related to the
significant time systems can remain active after their initial peak of
activity; if, after a near-Eddington peak in activity, the AGN
accretion rate declines in an approximate power-law fashion $\propto
(t/t_{0})^{-\beta}$, then for a system at much lower Eddington ratio
$\mdot$, it has been a time $\sim t_{0}\, \mdot^{-1/\beta}$ since the
peak of activity. For characteristic $t_{0}\sim10^{8}\,$yr suggested
by observational estimates of the AGN lifetime and $\beta\sim1-3$ from
a variety of different models \citep[see][and references
therein]{hopkins:seyferts,hopkins:mdot.dist},
this time is still small compared to the
Hubble time for high Eddington ratios $\mdot\gtrsim0.1$. The system is
therefore viewed with no significant evolution in its properties
between the peak of activity and the time at which it is observed.
However, at sufficiently low Eddington ratios $\mdot\ll 0.01$, the
initial trigger may have occurred at a much earlier time
$\gtrsim\,$Gyr (comparable to a Hubble time), implying that systems
observed at e.g.\ $z=0$ -- having decayed for Gyr down to these
Eddington ratios -- were active/forming at much earlier times, and
therefore are more biased than a system of similar mass first becoming
active at $z=0$. Modeling this in detail requires a detailed and
specific model for AGN lightcurves and triggering; we adopt the
results from \citet{mcquinn:helium.reionization.model} \citep[see][for
more details]{hopkins:clustering,hopkins:cusps.fp,hopkins:cusps.evol} 
to demonstrate the effects here;
however, many of the qualitative results are general.
That model uses identical assumptions to the model 
described here in \S~\ref{sec:mdot} (the same AGN luminosity 
function, light curves, Eddington ratio and host property distributions), 
but places the mock AGN in halos in a cosmological simulation 
according to the observationally constrained galaxy mass-halo 
mass relations \citep[see e.g.][and 
references therein]{moster:stellar.vs.halo.mass.to.z1}, and 
so can predict clustering and other properties. 

\begin{figure}
    \centering
    \scaleup
    \plotter{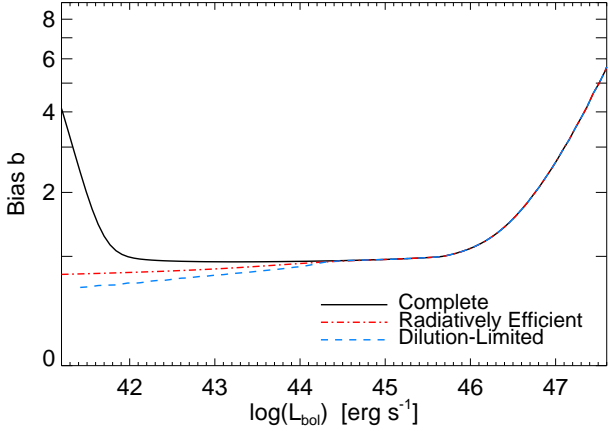}
    \caption{Dependence of clustering on luminosity at $z\sim0$ for
    the different samples described in Figure~\ref{fig:ledd.dist}. 
    At $L\gtrsim L_{\ast}$, the samples are all similar, with mostly 
    near-Eddington  systems; thus $L$ scales roughly with halo mass and there is a 
    steep $b(L)$. At lower $L$, the broad range of Eddington ratios leads to a 
    weak dependence of $b(L)$ and can cause differences in different samples
    \citep[see][]{lidz:clustering}. At very low $L$, the clustering 
    in a complete sample can increase again (depending on the exact lightcurve/lifetime 
    model), as the luminosity function can be dominated by 
    early-forming highly clustered BHs that have had time to 
    decay to very low luminosities. These low $\dot m$ systems would 
    not appear in radiatively efficient or dilution-limited samples.     
    \label{fig:clustering}}
\end{figure}

Figure~\ref{fig:clustering} shows the bias as a function of luminosity
taking into account the above effects and for our three different
``observational'' samples. At the bright end, there is no difference
between models and a reasonably steep dependence of clustering on
luminosity, because essentially all models agree that AGN must be
near-Eddington at these luminosities, implying that the luminosity
function is a sequence in BH/host mass. At luminosities $\lesssim
L_{\ast}$, however, the fact that non-trivial lightcurves allow the
population to be a mix of both high and low Eddington ratio systems
rather than a strict sequence in mass makes the dependence of
clustering amplitude on luminosity relatively weak, as is observed
\citep{porciani2004,
  adelbergersteidel:lifetimes,porciani:clustering,
  coil:agn.clustering,daangela:clustering} \citep[for more details regarding the modeling of clustering,
see][]{lidz:clustering}. For ``complete'' samples, the clustering
amplitude is predicted to {\em increase} at lower luminosities,
because of an increasing contribution of very massive systems that
have decayed to very faint luminosities (low Eddington ratios).
Indeed, observations of faint X-ray sources appear to support this,
with a surprisingly high clustering amplitude in several cases
\citep[see e.g.][]{constantin:llagn.clustering,
  plionis:xr.llagn.clustering,gilli08:xr.agn.clustering,
  hickox:multiwavelength.agn} -- higher
than observed in optical samples at the same redshift and similar or
higher luminosities \citep{croom:clustering,myers:clustering,daangela:clustering}. 

If, however, these sources are selected against in optical/IR samples,
then the clustering amplitude remains flat or decreases at low
luminosities, reflecting the fact that the most biased sources (which
are very massive, old galaxies that with AGN at very low Eddington
ratios) are excluded, leaving a sample dominated by lower-mass systems
at intermediate/high Eddington ratios.  This may explain why the
clustering amplitudes in X-ray and optical samples appear to disagree
at low luminosities.

\begin{figure}
    \centering
    \scaleup
    \plotter{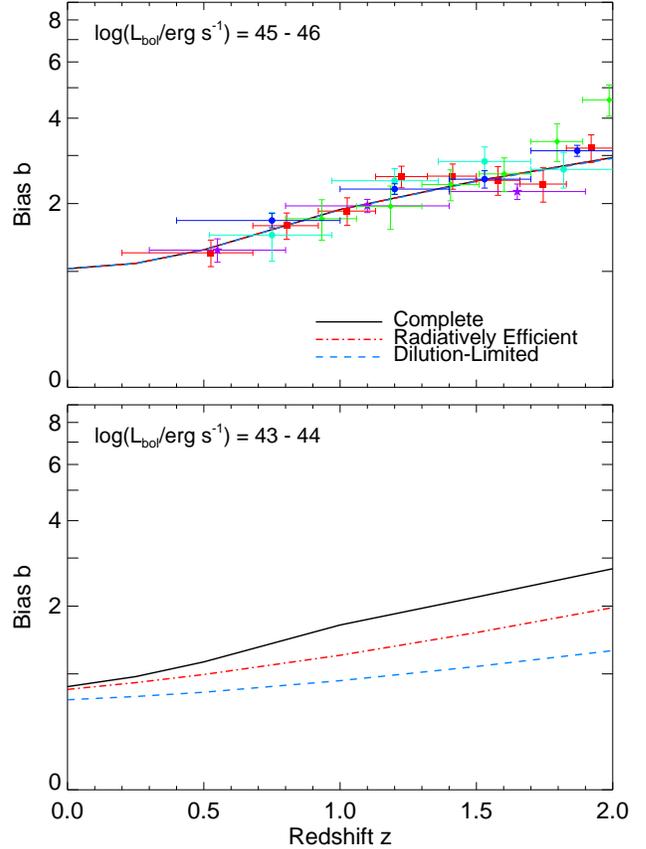}
    \caption{Dependence of AGN clustering on redshift in a
      intermediate-high luminosity sample ($\sim L_{\ast}$ quasars at
      these redshifts; {\em top}) and a low-luminosity sample
      (corresponding to $L_{2-10\,{\rm keV}}\sim10^{42}-10^{43}\,{\rm
        erg\,s^{-1}}$ or $M_{B}\sim-17$; {\em bottom}), taking into
      account the different selection effects described in
      Figure~\ref{fig:ledd.dist}.  For the $\sim L_{\ast}$ sample,
      where most observations have been made, the 
      different selection effects do not significantly change 
      the predicted bias, which is also consistent with the 
      observed   clustering of quasars of similar luminosities from optically
      selected samples in \citet[][red squares]{croom:clustering},
      \citet[][green diamonds]{porciani:clustering}, \citet[][cyan and
      blue circles]{myers:clustering,myers:clustering.old}, and
      \citet[][violet stars]{daangela:clustering}.  At very low
      luminosities, the effects described in
      Figure~\ref{fig:clustering} and \S~\ref{sec:morphology} lead to
      significant differences between samples; the 
      higher bias in X-ray samples has been observed in a number of 
      studies (see text). 
      \label{fig:clustering.z}}
\end{figure}

Figure~\ref{fig:clustering.z} shows the clustering amplitude expected
as a function of redshift for our standard three samples of AGN at two
different intervals in bolometric luminosity. At a relatively high
luminosity (top panel), which corresponds roughly to $\sim L_{\ast}$
in the AGN luminosity function over the redshift range $z\sim0.5-2$
\citep[see e.g.][]{hopkins:bol.qlf}, the different selection methods
yield nearly identical clustering amplitudes, in agreement with a
variety of observations.
At low luminosities, however ($L_{\rm bol}\sim10^{43}-10^{44}\,{\rm
  erg\,s^{-1}}$, corresponding to $L_{2-10\,{\rm
    keV}}\sim10^{42}-10^{43}\,{\rm erg\,s^{-1}}$), the bias versus
redshift depends on the selection criteria (bottom panel in
Fig.~\ref{fig:clustering.z}). In particular, at these luminosities,
AGN selected in dilution-limited samples are preferentially at high
Eddington ratio and therefore, smaller BHs in low-mass,
weakly-clustered hosts.  The bias as a function of redshift is
somewhat sensitive to the exact model we adopt, but the selection
effects are in any case clearly not negligible -- they can lead to a
factor of $\sim2$ difference in bias, which corresponds to up to an
order-of-magnitude difference in the implied host halo mass at these
redshifts and bias values. This is qualitatively similar to what is
observed in radio versus X-ray versus infrared-selected AGN samples
(e.g., \citealt{hickox:multiwavelength.agn}).

\section{Differences in Host Masses and Morphologies}
\label{sec:morphology}

At a given luminosity, different Eddington ratios imply different BH
masses and thus different host masses. Given the strong correlation
between galaxy morphology and stellar mass, this naturally leads to
variations in host morphologies found in different samples.  
Recall, our model includes both the host total mass and bulge/disk 
mass distributions. For convenience, we therefore 
adopt the bulge-to-total ratio $B/T$ as our morphological
proxy of interest.

Some caution is warranted here -- making such a conversion implicitly
assumes that whatever fuels objects is statistically random at each
bulge mass/luminosity, and does not select out a specific morphological
sub-population at a given stellar mass.  This is not necessarily true
if e.g.\ mergers or disk instabilities are the dominant means of AGN
fueling. Nevertheless, some samples suggest that, at least at low
luminosities where fueling an AGN requires only a small gas supply,
the host distribution spans a wide range and does not show a strong
preference for specific morphological types \citep{bahcall:qso.hosts,
  malkan:qso.host.morph,heckman:local.mbh,
  falomo:qso.hosts,sanchez:qso.host.colors,
  rigby:qso.hosts,nandra:qso.host.colors}.  At higher luminosities
this is probably not the case, but as before, it is the low-luminosity
regime where the Eddington ratio distribution is broad and has a
significant effect.

\begin{figure}
    \centering
    \scaleup
    \plotter{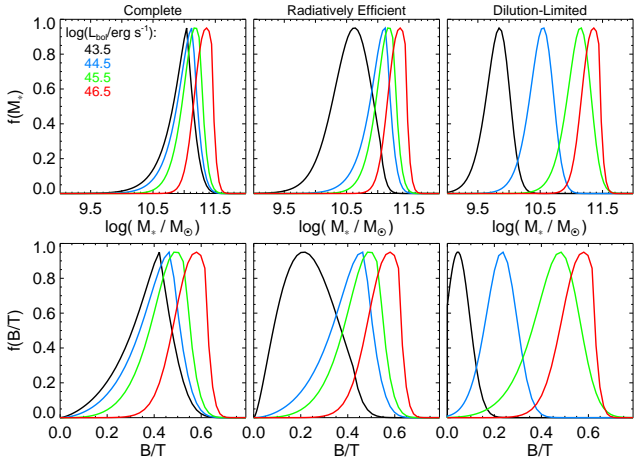}
    \caption{Model distribution of host galaxy stellar masses and B/T
      (bulge-to-total stellar mass ratio, our proxy for morphology) as
      a function of bolometric luminosity for the different selection
      samples considered in Figure~\ref{fig:ledd.dist}. In a complete sample
      (X-ray or narrow line), even low-$L$ AGN populations are
      dominated by relatively massive ($M\sim
      10^{10.5}-10^{11}\,\msun$) early-type (Sa/S0/E) hosts, as
     is observed \citep[e.g.][]{kauffmann:qso.hosts}.  In a radiatively
      efficient or dilution limited sample, however, the 
      requirement of moderate Eddington ratios at low $L$ leads to
      host masses and morphologies that are more luminosity-dependent: lower-$L$ AGN in these
      samples will be hosted by lower-mass, late-type
      ($B/T\lesssim0.3$) galaxies. This can explain the 
      tension between the ``conventional wisdom'' that 
      Seyferts live in disks (from optical-IR samples) and the contradictory results
      from recent narrow-line or X-ray samples that 
      moderate-luminosity AGN have primarily early-type hosts.
      \label{fig:morphology}}
\end{figure}

The predicted host stellar mass and morphology distributions are shown
in Figure~\ref{fig:morphology}. For the reasons discussed in
\S~\ref{sec:clustering}, a sample that includes objects of all types
(including those that are radiatively inefficient or diluted in the
optical/IR) is dominated by relatively massive ($\sim L_{\ast}$)
early-type galaxies at all luminosities. This conclusion has been reached 
observationally in e.g.\ narrow line and deep X-ray samples: 
\citet{kauffmann:qso.hosts}, 
\citet{kewley:agn.host.sf} 
and \citet{hao:local.lf} show that host luminosity and mass 
are nearly independent of AGN luminosity, with early-type morphologies 
predominant (similar to the X-ray conclusions discussed in 
Figure~\ref{fig:ledd.dist.cum}). 
However, in a sample with an
implicit Eddington ratio selection, only low-mass BHs are present in
the sample at low luminosities, which leads to a host sample dominated
by less massive, more disk-dominated and star-forming host galaxies.

\begin{figure}
    \centering
    \scaleup
    \plotter{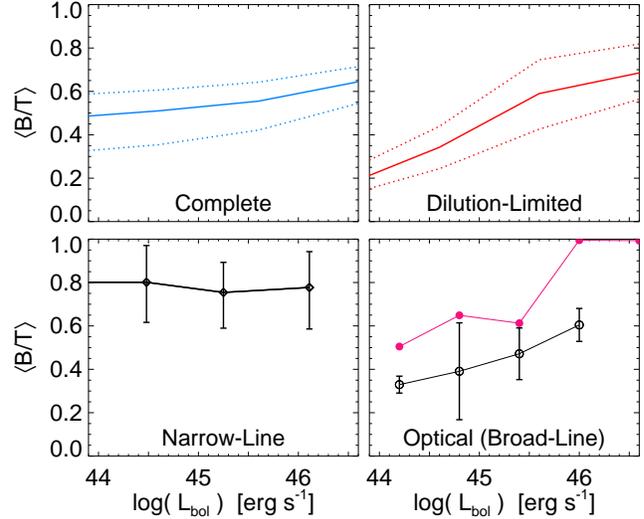}
    \caption{{\em Top:} Median bulge-to-total ratio 
    as a function of luminosity, 
    for the complete ({\em left}) and dilution-limited ({\em right}) 
    samples in the model. Dotted lines show the 
    $\pm1\,\sigma$ range in the population. {\em Bottom Left:} 
    Corresponding observational 
    estimates from the narrow-line SDSS AGN sample  \citep{kauffmann:qso.hosts}, 
    converting $L_{O\,III}$ to bolometric and concentrations to $B/T$ given the mean 
    observed relation in \citet{graham:bulges} (error bars 
    are the dispersion in the population, not errors in the median). 
    {\em Bottom Right:} Same, based on the optically-selected PG quasar  
    sample from \citet{dunlop:qso.hosts}. We show the results both for their 
    sample with ``robust'' morphologies (reliable bulge/disk 
    decomposition; black circles) and their full sample (magenta squares). 
    \label{fig:host.morph.obs}}
\end{figure}

This result can potentially resolve long-standing discrepancies
between different observations of AGN host galaxies, in samples that
should select for similar AGN bolometric luminosities. We illustrate
this in Figure~\ref{fig:host.morph.obs}, where we plot the median
$B/T$ as a function of AGN luminosity for two model samples: our
complete sample and the dilution limited/optical broad-line sample 
(we do not plot host masses/luminosities because this information is 
already effectively presented in Figure~\ref{fig:ledd.dist.cum}, given 
$M_{\rm BH}\propto M_{\rm bulge}$; moreover the host morphology 
distribution is the primary focus of many studies attempting to 
constrain AGN fueling processes). We
compare with roughly comparable observational samples: the deep
narrow-line sample of \citet{kauffmann:qso.hosts} (which includes
faint objects with luminosities well below those of their hosts) and
the broad-line PG quasar sample (all of which have $L_{\rm AGN}\gtrsim
L_{\rm host}$) from \citet{dunlop:qso.hosts} and
\citet{floyd:qso.hosts}.\footnote{Note that the fitting procedures for
  $B/T$ in these samples differ from the calibration in
  \citet{balcells:bulge.scaling} that we use for our predictions --
  therefore some systematic offsets are expected. The qualitative
  trends with mass are of primary interest here.}

The ``conventional wisdom,'' derived from early faint AGN samples that
were dominated by broad-band optical/UV color selection, has been that
Seyferts reside predominantly in disk-dominated, low-mass galaxies
\citep[][]{adams:seyfert.host.imaging,
  smith:radio.loud.quiet.agn.hosts, granato:seyfert.host.morph,
  mcleod:seyfert.host.morph}; this is still true in modern samples
that adopt similar selection methods, i.e., that require a moderate
UV/optical/IR luminosity from the AGN relative to the host galaxy
\citep[see e.g.][and references therein]{martini:seyfert.host.morph,
  peletier:seyfert.morph.imaging,dunlop:qso.hosts,
  zakamska:qso.hosts}.  These inferences are consistent with our
models for a sample with an effective minimum Eddington ratio.
However, sufficiently deep narrow-line and X-ray surveys (``complete''
surveys) find instead that the AGN host population is dominated by
massive early-type galaxies, at all luminosities.  At a given
(similar) bolometric luminosity, these surveys suggest that the AGN
host fraction rises steeply as a function of bulge mass, in contrast
to the results from Type 1-sensitive surveys.  As
Figure~\ref{fig:host.morph.obs} shows, both results follow naturally
once one takes into account a broad intrinsic distribution of
Eddington ratios and the observational selection effects highlighted
throughout this paper.

\section{Discussion and Conclusions}
\label{sec:discussion}

We have demonstrated that, given a non-trivial Eddington ratio
distribution, two simple effects will give rise to potentially
important differences in the AGN populations selected by different
observational criteria.  Even if the physics of accretion were
independent of $\dot m$, an AGN will inevitably be much less luminous
than its host galaxy in a given band at sufficiently low Eddington
ratio.  Dilution is, of course, inevitable, and in most surveys AGN
will be classified as ``normal galaxies'' if the AGN luminosity is
less than that of the host. In the optical/IR, this happens at
surprisingly high Eddington ratios, $\mdot\sim0.01-0.1$, for most
low-luminosity populations, which is why the effects of dilution can
have such a significant impact on AGN surveys.  Moreover, at low
accretion rates $\mdot\lesssim0.01$, AGN may transition to a
radiatively inefficient state, in which the optical and infrared
emission decline rapidly, and eventually the broad-line region may
disappear entirely (see footnote 2 in \S \ref{sec:intro} for more
discussion of how the broad line emission disappears).
Regardless of the specific spectral changes that may occur, the
phenomenon of low-efficiency accretion is increasingly
well-established and at low accretion rates populations of AGN with
little to no broad-line emission appear \citep{ho:llagn.seds,
  moran:hblr.seyferts,tran:hblr.2,
  nicastro:hblr,brightman:non.hblr.check,bianchi:nonhblr}.

At high luminosities, observed systems are uniformly at relatively
high Eddington ratios, and so it is unlikely that either 
dilution or radiative
inefficiency is a significant issue.  At low luminosities,
however, the ``true'' Eddington ratio distribution of all AGN, as
constrained by sufficiently deep narrow-line, X-ray, and radio
surveys, is broad and extends to lower $\mdot$ with decreasing
bolometric luminosity \citep[roughly as $\lambda = L_{\rm bol}/L_{\rm
  Edd} \propto L_{\rm bol}^{0.5-1}$ at sufficiently low luminosities;
see, e.g.,][]{fiore:type2.lx.vs.lhost,
    silverman:hx.spacedensity.ldde,
  hickox:multiwavelength.agn}; this is
consistent with some theoretical models of AGN lightcurve evolution
\citep{hopkins:mdot.dist}.

Given the selection effects implicit in the construction of a
broad-line or Type 1 optical/mid-IR AGN sample, our results imply that
dilution and radiative efficiency lead to these samples preferentially
sampling the high-$\mdot$ end of the population, with
$\mdot\gtrsim0.1$ and a relatively narrow range in $\mdot$ even at low
luminosities.  This is indeed observed
(Figures~\ref{fig:ledd.dist}-\ref{fig:ledd.dist.cum}).  The strong
impact of selection criteria is illustrated by the distinct Eddington
ratio distributions recovered in X-ray samples (where dilution is
weak, and radiatively efficiency declines less rapidly) as compared to
optical samples \citep[compare][]{greene:active.mf}.

For the same bolometric luminosity, different Eddington ratios imply
different BH masses and, correspondingly, host galaxy
masses. Low-luminosity AGN hosts in sufficiently deep narrow
line/X-ray/``complete'' samples should cluster around a narrow range
in host mass, preferentially massive ($\sim L_{\ast}$), early-type
galaxy hosts, as observed \citep{kauffmann:qso.hosts,
hao:local.lf.selection,hao:local.lf,kewley:agn.host.sf,
silverman:qso.hosts,hickox:multiwavelength.agn}.  However,
low-luminosity AGN in dilution (or radiative efficiency)-limited samples
will often be found in less massive hosts, which are correspondingly
less bulge-dominated
(Figures~\ref{fig:morphology}-\ref{fig:host.morph.obs}).  This is also
observed, and can reconcile conflicting claims in the literature
regarding the hosts of Seyfert galaxies: samples based on optical/UV
excess selection have traditionally found that these objects reside in
less massive late-type disks \citep{depropris:seyfert.host.morph,
  martini:seyfert.host.morph,knapen:seyfert.morphology,dunlop:qso.hosts,
  rigby:qso.hosts} while more recent X-ray or
narrow line surveys find that AGN preferentially lie in massive early
type hosts (references above).  As it is well-established that
galaxies of different masses and morphologies cluster differently,
these same effects also lead to differences in the expected clustering
(Figure~\ref{fig:clustering}): faint X-ray/narrow-line/``complete''
AGN samples should exhibit higher clustering amplitude than
corresponding optical samples. This is consistently observed when
comparing X-ray and optical/IR AGN clustering properties at low
luminosities \citep[see e.g.][and references
therein]{porciani:clustering,
  myers:clustering,
  plionis:xr.llagn.clustering}.

In most surveys, AGN are identified as ``obscured'' or ``unobscured''
in the optical based on a simple classification, namely whether the
optical spectrum is dominated by a broad-line AGN SED or galaxy light;
we have shown, however, that both dilution and radiative inefficiency
will lead to the classification of an AGN as obscured (e.g.,
Fig. \ref{fig:compare.templates}). The natural expectation is that the
fraction of sources which are optically dim for these reasons should
increase with decreasing luminosity. It is in fact traditionally
inferred that the intrinsically ``obscured'' (non Type 1) AGN fraction
behaves in just this manner \citep{hill96:sf.abs.in.radio.gal,
  willott00:optical.qso.frac.vs.l,
  steffen03:agn.obsc.vs.l.z,ueda03:qlf,grimes04:obsc.frac.vs.l,
  hasinger04:obsc.frac.and.xrb,sazonov.revnivtsev04:local.hx.qlf,barger:qlf,
  gilli:obscured.fractions,
  hickox:bootes.obscured.agn,hasinger:absorption.update}. We find,
however, that the fraction of AGN that are diluted (and may also be
radiatively inefficient) scales in a similar manner to the observed
optically ``obscured'' fraction, with both luminosity and
redshift. The effect is not large enough to account for all optically
dim objects (in particular at high L), but the fraction of ``true''
obscured sources could, in principle, be as low as $\sim20-40\%$, with
a much weaker dependence on both luminosity and redshift than that
traditionally inferred (see
Figures~\ref{fig:obscuration}-\ref{fig:obscuration.hardness}).

Using X-ray hardness as a measure of obscuration represents an
improvement over the ``optically dim'' criterion in many ways (e.g.,
it should not be affected by simple dilution).  However, the expected
X-ray SEDs of AGN in radiatively inefficient states are relatively
hard power-laws for $\dot m \sim 10^{-2}$ (Figure
\ref{fig:riaf.templates}).  They are sufficiently hard, in fact, that
the systems would be classified as obscured/hard sources by the
standard observational criterion -- which uses a simple hardness ratio
to probe the presence or absence of an absorption cutoff, given the
lack of full X-ray spectra at the signal to noise ratios available for
most high redshift sources. Convolving with the observed Eddington
ratio distribution, we find that RIAFs could explain a large fraction
of the X-ray ``obscured'' sources at low luminosities ($L_{\rm
  bol}\lesssim10^{45}\,{\rm erg\,s^{-1}}$, or $L_{2-10\,{\rm
    keV}}\lesssim3\times10^{43}\,{\rm erg\,s^{-1}}$). In particular,
just as with the optical samples,
we find that the ``true'' obscured fraction could be as low as $\sim
20-40\%$, with a much weaker dependence on luminosity and/or redshift
than has been suggested by observations based on hardness ratio
distributions.  It is important to stress that we are not questioning
the fact that {\it local} samples of Seyferts indicate a large
population of obscured sources.  Rather we are pointing out that the
hard X-ray spectra detected in higher redshift samples may in many
cases be {\it intrinsic} (having to do with accretion physics), rather
than indicative of absorption.

Consistent with the conclusions drawn here, detailed observations have
revealed that dilution accounts for a significant fraction of the
optically dim population at luminosities $L_{\rm
  bol}\lesssim10^{45}\,{\rm erg\,s^{-1}}$
\citep{moran:dilution.making.seyfert2s,hao:local.lf,
  severgnini:xmm.diluted.faint.agn,
  georgantopoulos:xbong.dilution,georgakakis:faint.agn.dilution,
  eckart:low.l.agn.dilution,
  garcet:optical.xray.agn.classes,
  gruppioni:mir.followup.optical.agn.classes}.  Specifically, a large
fraction of the sources identified as Type 2 on the basis of a lack of
observed broad lines
or optical/IR AGN continuum do not show similar evidence for
obscuration based on X-ray followup \citep[see
e.g.][]{akylas:obscured.fractions.vs.lx,
caccianiga:true.obscured.agn.fracs.xbs.dilution,
hasinger:absorption.update}. 
In many optical/IR studies the Type 2 fraction remains as high as
$f_{\rm obscured}\sim0.6-0.7$ up to luminosities $L_{\rm
  bol}\lesssim3\times10^{45}\,{\rm erg\,s^{-1}}$ ($L_{2-10\,{\rm keV}}
\lesssim10^{44}\,{\rm erg\,s^{-1}}$), by which point the obscured
fraction suggested by X-ray followup has dropped to $f_{\rm
  obscured}\sim0.2$ 
  \citep[see also][]{tajer:xr.optical.obscured.agn}; 
  \citet{della.ceca:true.obscured.agn.fracs} and 
  \citet{caccianiga:dilution.dominates.low.l.agn.obsc}, for example,  
  find that this corresponds closely to what is seen in deeper follow-up 
  of X-ray selected and apparently optically dim sources in the XMM-Newton 
  survey. 
\citet{hao:local.lf} estimate from a narrow-line (SDSS galaxy) sample 
that the true obscured 
fraction remains a constant $\sim30-40\%$ from luminosities 
$L_{\rm bol}\sim10^{42}-10^{45}\,{\rm erg\,s^{-1}}$ 
($L_{O\,{\rm III}}\sim 10^{5}-10^{8}\,L_{\sun}$).   
  Thus $\sim 50-80 \%$ of the claimed obscured
sources in this luminosity range (above $L_{\rm bol}\sim10^{44}\,{\rm
  erg\,s^{-1}}$ or $L_{2-10\,{\rm keV}}\sim 10^{42}\,{\rm
  erg\,s^{-1}}$) appear to simply be diluted.

Note that, although we have found that the obscured fraction could be
as low as $\sim20\%$, we are not strongly advocating this specific
number; this low constant level of obscuration (relatively independent
of luminosity and redshift) corresponds to the maximum effect of
dilution on observational surveys, given the uncertainties in our
approach and the observations.  The observational studies above
suggest that the true baseline obscured fraction may be $\sim30-40\%$
or even slightly higher.  Our key point is that the observed {\em
  trend} of obscuration with luminosity is very sensitive to dilution.
%
And although it has been well appreciated that
dilution can significantly affect low luminosity AGN samples selected
in the optical/IR, it has not been as widely appreciated that it also
biases the inferred Eddington ratios, host mass and morphology
distributions, and even clustering amplitudes (as we have
demonstrated). Moreover it is not clear whether any trivial
``correction'' can be applied to these samples to correct for these
effects.

More detailed observations probing the type of systems we appeal to
here are clearly important.  Unfortunately, they are also
difficult. As noted above, the sources of interest -- those that may
dominate the ``obscured''/hard population at low luminosities in large
X-ray surveys -- are {\em not} analogous to well-studied local Seyfert
2 galaxies (which are indeed obscured by dust and gas).  The latter
are primarily nearby, relatively active systems in low-mass,
star-forming and/or disky host galaxies. In contrast, the optically
dim/hard population that dominates at low luminosities in large X-ray
surveys contains a large fraction of massive elliptical galaxies
($M_{\ast}\gtrsim10^{11}\,\msun$, and BH mass
$\sim10^{8}-10^{9}\,\msun$) firmly in the ``red sequence'' (i.e.\
without star formation or significant gas content), with low Eddington
ratios $\lambda\sim10^{-3}-10^{-2}$, and relatively high clustering
amplitudes \citep[see e.g.\
Figures~\ref{fig:ledd.dist.cum}-\ref{fig:host.morph.obs} and]
[]{fiore:type2.lx.vs.lhost,kauffmann:qso.hosts,brand:red.gas.xr.sources,
  nandra:qso.host.colors,silverman:qso.hosts,hickox:multiwavelength.agn}.
In some ways this population is more analogous to radio galaxies (but
many of the systems of interest are still radio-quiet), for which a
number of models and observations suggest that low Eddington ratios
and differences in accretion physics (relative to local Seyferts) are
important \citep{smith:radio.loud.quiet.agn.hosts,
  falcke:radio.vs.mdot,meier:jets.in.adaf,ho:radio.vs.mdot,
  maccarone:agn.riaf.connection,best:radio.loudness,
  fender:radio.mdot.review,merloni:synthesis.model}.

A $L_{\rm bol}\sim10^{44}\,{\rm erg\,s^{-1}}$ BH in the radiatively
inefficient regime (with $\lambda = L_{\rm bol}/L_{\rm Edd}\sim0.001$
for $\mdot<0.01$) implies a $\gtrsim10^{9}\,\msun$ BH, and there are
only a few such objects within $\sim100$\,Mpc. Interestingly, of the
six such objects in the local volume with well-measured BH masses
\citep[from][]{tremaine:msigma,marconihunt,haringrix}, at least four
(M87, NGC 1399, NGC 4649, and IC 1459) are considered RIAF candidates
\citep[albeit with some debate; and in e.g.\ the case of M87 the
presence of a jet complicates comparisons;][]{sambruna:asca.xr.agn,
  dimatteo:riaf.candidates,allen:riaf.candidates,sulkanen:adaf.doublecheck,
  fabbiano:1459}; they have luminosities $L_{X}\sim
10^{40}-10^{43}\,{\rm erg\,s^{-1}}$ ($L_{\rm
  bol}\sim10^{41}-3\times10^{44}\,{\rm erg\,s^{-1}}$) and extremely
hard, power law-like X-ray SEDs (effective photon indices $\Gamma =
0.7-1.4$).  These systems are distinct from many better-studied nearby
Seyferts in low-mass galaxies, which have softer X-ray spectra
$\Gamma\sim2$ and Eddington ratios $\sim10^{-2}$.  These are small
number statistics, but we reach similar conclusions if we consider the
$\sim$ dozen nearby BHs down to $\sim$few $\times10^{8}\,\msun$
\citep[see][]{evans:6251,soria:adaf.candidate.xr.obs,soria:adaf.candidates.stellarwinds,
  ellis:lum.compilations,gonzalez:liner.nature}.  Generalizing from
this very small sample, if $\sim1/2$ of $M_{\rm
  BH}\gtrsim10^{9}\,\msun$ BHs are active in this luminosity range
with X-ray SEDs as hard as observed in this subset, this is already
sufficient to account for $>50\%$ of the observed hard X-ray
(apparently ``obscured'') population at these luminosities in large
samples from $z\sim0-1$. But deep studies searching for e.g.\ polarized 
thin-disk signatures could set clear lower limits on the fraction 
of such sources with ``traditional'' obscuration. Clearly, more detailed observations of the
(admittedly small number of) local analogues is critical to inform our
speculation that these systems may contribute significantly in large
surveys.

There has been considerable debate about the exact physics dominant at
low Eddington ratios (e.g., an ADAF, a jet-dominated spectrum, or some
other RIAF model; see, e.g., \citealt{guainazzi:1052,
  angelini:lmxb.hx.em,loewenstein:lmxb.hx.contrib,
  lumsden:scattered.halpha.seyfert2s,gallo:low.abs.type2s,
  gliozzi:naked.agn,ghosh:all.obscuration,
  brightman:non.hblr.check}). We stress that we are not attempting to
advocate any {\em specific} model among these.  Rather, the key
question is to test is whether the lack of optical emission and the
hard X-ray spectra are intrinsic or the result of a large column, 
or even whether or not the X-ray data supports the optical conclusions 
(or whether those can be entirely accounted for by dilution).  To
do so, more observations to break the degeneracies between accretion
state, dilution, and genuine obscuration are clearly important.
Ideally, high-resolution nuclear optical and X-ray spectroscopy and
optical polarimetry could be applied, as in the most well-studied
local Seyferts, to break the degeneracies between these possibilities
and the most popular alternatives (``receding torus'' models, in which
a circumnuclear torus leads to geometric obscuration at all
luminosities, but with a luminosity-dependent covering angle).
Although these observational tests may be possible for the local
analogues noted above, most systems at these low luminosities in large
surveys cannot be studied in such detail. Observations must instead
rely on indirect probes; these probes do, however, allow statistical tests that
are not possible in the small local populations, and the application of integral arguments 
following BH demographics \citep[see e.g.][]{brand:red.gas.xr.sources}. 

Perhaps the largest theoretical uncertainty in estimating the
importance of radiatively inefficient accretion for large AGN samples
is the rapidity of the transition from a thin disk to a RIAF.  If the
luminosity declines too rapidly in all bands, say, then all objects
below some $\dot m$ will simply drop out of all samples, and will not
introduce significant differences between AGN samples in various
wavelengths. If, on the other hand, the thin/thick transition occurs
very slowly with decreasing $\dot m$, then the SEDs will remain
relatively X-ray soft and optically/IR bright at lower luminosities
(Fig. \ref{fig:riaf.templates}), leading to less of a contribution
from such sources to the X-ray hard/optically dim
populations. Constraints on the contribution of such sources to faint
``obscured'' AGN samples, therefore, are a potentially new and
powerful way to determine in detail how (and indeed whether or not) 
AGN behavior changes at low accretion rates.

If dilution and radiative inefficiency are indeed significant at the
levels estimated here, they can naturally account for observed
differences in AGN populations selected at various wavelengths and via
different techniques.  The apparently contradictory host galaxy
properties, clustering, and Eddington ratio distributions from
different samples can, in fact, be reconciled and understood using a
coherent theoretical framework.
The importance these effects for large AGN
samples is also intimately related to the problem of AGN fueling.
Black holes that gained most of their mass in early, high-accretion
rate episodes and have since decayed to low luminosities over a Hubble
time -- typical in merger-driven fueling scenarios -- may show up as
low-$\mdot$ populations in massive, early-type hosts. On the other
hand, systems that continuously or stochastically fueled -- expected
in scenarios where BHs are fueled at lower levels by e.g.\ stellar
winds or bars in disk galaxies -- will plausibly show up as
higher-$\mdot$ populations in intermediate mass, later-type hosts.

The results highlighted here also represent an important revision and
caveat to the unified model of AGN.  Even if AGN physics are perfectly
self-similar, dilution implies that objects selected in optical,
infrared, or X-ray samples are {\em not} random subsets of the same
parent population (with just viewing angle differences).  Rather,
dilution leads to host galaxy-dependent selection effects, biasing
e.g.\ the Eddington ratio distribution, host
luminosity/mass/age/clustering distributions, and other
properties. Moreover, radiatively inefficient populations could
represent a large fraction of the ``obscured'' population at low
luminosities.  This should not be viewed as a limitation of these
observations -- rather, it presents the opportunity for
multiwavelength comparisons to develop new constraints on AGN fueling
mechanisms, feedback, and host populations.  Since AGN selection may
not be strictly independent of host properties at low luminosities,
comparison of the demographics selected in different wavelengths can
break degeneracies between different fueling and host evolution
models.

\acknowledgements
We thank Paul Martini and Smita Mathur for helpful
discussions. We also thank the anonymous referee for 
valuable comments and suggestions. This work
was supported in part by NSF grants ACI 96-19019, AST 00-71019, AST
02-06299, and AST 03-07690, and NASA ATP grants NAG5-12140,
NAG5-13292, and NAG5-13381.  Support for PFH was provided by the
Miller Institute for Basic Research in Science, University of
California Berkeley.  E.Q. is supported in part by NASA grant
NNG06GI68G and the David and Lucile Packard Foundation.
\\

\bibliography{/Users/phopkins/Documents/lars_galaxies/papers/ms}

\end{document}